\documentclass[draft]{agujournal2019}
\usepackage{url} 
\usepackage{lineno}
\usepackage[inline]{trackchanges} 
\usepackage{soul}
\usepackage{amsmath,mathtools}
\usepackage{txfonts} 
\usepackage[utf8]{inputenc}
\usepackage[T1]{fontenc}

\usepackage{wasysym} 

\usepackage{rotating}
\usepackage{amssymb}    
\usepackage{cases}  
\usepackage{multirow} 
\usepackage{xcolor} 

\DeclareRobustCommand*{\deg}{\mathrm{^\circ}}

\DeclareMathOperator{\tr}{tr}

\usepackage{xfrac}
\usepackage{etoolbox}
 
\newcommand{\qty}[2]{$#1\,\textnormal{#2}$}

\usepackage{verbatimbox}
\usepackage{cprotect}

\renewcommand{\v}[1]{\ensuremath{\mathbf{#1}}} 
 
\let\baraccent=\= 
\renewcommand{\=}[1]{\stackrel{#1}{=}} 

\draftfalse

\journalname{JGR: Space Physics}

\begin{document}

%
%

\title{Making waves: Mirror Mode structures around Mars observed by the MAVEN spacecraft}

%
%




\authors{Cyril Simon Wedlund\affil{1}, Martin Volwerk\affil{1}, Christian Mazelle\affil{2}, Jasper Halekas\affil{3}, Diana Rojas-Castillo\affil{4}, Jared Espley\affil{5}, and Christian Möstl\affil{1}}


\affiliation{1}{Space Research Institute, Austrian Academy of Sciences, Graz, Austria}
\affiliation{2}{Institut de Recherche en Astrophysique et Planétologie (IRAP), Université de Toulouse, CNRS, UPS, CNES, Toulouse, France}
\affiliation{3}{Department of Physics and Astronomy, University of Iowa, Iowa City, IA, USA}
\affiliation{4}{Instituto de Geofísica, Universidad Nacional Autónoma de México, Coyoacán, Mexico} 
\affiliation{5}{NASA Goddard Space Flight Center, Laboratory for Planetary Magnetospheres, Greenbelt, MD, USA}

%



\correspondingauthor{C. Simon Wedlund}{cyril.simon.wedlund@gmail.com}




\begin{keypoints}
\item We detect mirror mode structures at Mars, about 15-30 solar wind thermal proton gyroradii in size.
\item For the first time, we use MAVEN magnetic field, ion and electron data to characterise them fully.
\item Located in the deep magnetosheath, they are likely created behind the quasi-perpendicular shock.
\end{keypoints}

%
%

%
%


\begin{abstract}
We present an in-depth analysis of a time interval when quasi-linear mirror mode structures were detected by magnetic field and plasma measurements as observed by the NASA/Mars Atmosphere and Volatile EvolutioN (MAVEN) spacecraft. We employ ion and electron spectrometers in tandem to support the magnetic field measurements and confirm that the signatures are indeed mirror modes. Wedged against the magnetic pile-up boundary, the low-frequency signatures last on average $\sim 10$\,s with corresponding sizes of the order of 15--30 upstream solar wind proton thermal gyroradii, or 10--20 proton gyroradii in the immediate wake of the quasi-perpendicular bow shock. Their peak-to-peak amplitudes are of the order of 30--35\,nT with respect to the background field, and appear as a mixture of dips and peaks, suggesting that they may have been at different stages in their evolution. Situated in a marginally stable plasma with $\beta_{||}\sim 1$, we hypothesise that these so-called magnetic bottles, containing a relatively higher energy and denser ion population with respect to the background plasma, are formed upstream of the spacecraft behind the quasi-perpendicular shock. These signatures are very reminiscent of magnetic bottles found at other unmagnetised objects such as Venus and comets, also interpreted as mirror modes. Our case study constitutes the first unmistakable identification and characterisation of mirror modes at Mars from the joint points of view of magnetic field, electron and ion measurements. Up until now, the lack of high-temporal resolution plasma measurements has prevented such an in-depth study.
\end{abstract}

%
%

%


%
%
%
%

\section{Introduction}
	Mirror mode-like (MM) structures have been found everywhere in solar system plasmas \cite{tsurutani_magnetosheath_2011}, from the solar wind \cite{kaufmann_large-amplitude_1970,winterhalter_magnetic_1995,bale_magnetic_2009} to Earth \cite{tsurutani_drift_1984,lucek_mirror_1999}, Mars \cite{bertucci_mgs_2004,espley_observations_2004} and Venus \cite{volwerk_first_2008,volwerk_mirror-mode-like_2008}, Jupiter \cite{erdos_statistical_1996} and Saturn \cite{violante_observations_1995}, as well as comets \cite{mazelle_ultra_1991,glassmeier_mirror_1993,volwerk_mass-loading_2016}. MM waves are low-frequency long-wavelength transverse waves, usually linearly polarised and non-propagating in the plasma rest frame. Their non-linear evolution has been discussed both for electrons and protons \cite{kivelson_mirror_1996,balikhin_mirror_2010,soucek_cluster_2011}. An MM structure classically takes the spatial shape of a magnetic bottle imprisoning pockets of high-density plasma drifting with the ambient plasma. Thus, in the data, MMs commonly appear as sudden dips or peaks in the magnetic field intensity, anti-correlated with plasma density variations and with only little magnetic field angular variation across the structure \cite{tsurutani_magnetosheath_2011}. In competition with Alfvén ion cyclotron waves, with which they are co-generated in the plasma, MMs typically grow in a high-$\beta$ plasma ($\beta_{||}>1$) from an ion temperature anisotropy, itself triggered by any asymmetry upstream in the solar wind flow \cite{gary_mirror_1992}. The anisotropy either arises from the ring-beam distribution of locally new picked-up ions, as seen at Mars, Venus or comets, or as the downstream result of a quasi-perpendicular bow shock crossing, as seen on all objects, magnetised or unmagnetised \cite{tsurutani_mirror_2011}. The mirror mode instability criterion (MMIC) is usually expressed as \cite{hasegawa_drift_1969}:
\begin{linenomath}
\begin{equation}
		{\rm MMIC} = 1 + \sum_i{ \beta_{i\perp} \left( 1 - \frac{T_{i\perp}}{T_{i\parallel}}\right)} < 0
		\label{eq:instabilityCriterion}
	\end{equation}
\end{linenomath}
where $\parallel$ and $\perp$ denote the directions parallel and perpendicular to the ambient magnetic field direction. $T_i$ is the ion temperature and $\beta_{i\perp} = 2\mu_0\,N_i k_B T_{i\perp}/|B|^2$ the perpendicular plasma beta with ion density $N_i$. In Equation\,(\ref{eq:instabilityCriterion}) the sum is over all ion species $i$ although electrons, neglected here, should rigorously speaking also be included as they contribute to fulfilling the instability criterion, although their anisotropy is usually smaller. MM structures are often organised in trains and convected away at the ambient plasma velocity (so-called drift-mirror instability). They partake in the local plasma dynamics of the magnetosheath by limiting, together with the co-generated Alfvén ion cyclotron instability, the temperature anisotropy in the plasma \cite{soucek_properties_2008,Soucek2015}. Moreover, in the magnetosheath the presence of heavy ions, such as alpha particles from the solar wind or locally generated heavier ions such as O$^+$, tends to favour the growth of MMs over the Alfvén ion cyclotron mode. This results in MMs dominating in the observations \cite<see>[for simulation work and references to observations therein]{shoji_mirror_2009}.

	At Venus, MM-like magnetosheath structures typically last from a few seconds up to $30$\,s as observed by Venus Express (VEX) on its $24$-h orbit around the planet \cite{volwerk_mirror-mode-like_2008,volwerk_mirror_2016}. This suggests that these structures may become as large as one planetary radius (assuming a plasma speed of about \qty{200}{km/s} in the magnetosheath). 
	\citeA{bader_proton_2019}, using VEX-ASPERA-4 ion mass analyser data \cite{barabash_analyser_2007}, showed that the slight ion temperature anisotropy arising at the bow shock was likely responsible for the generation of proton cyclotron and MM waves. Perpendicular heating of the local plasma was found to be most prominent in the subsolar magnetosheath region, coinciding with MM-unstable conditions as calculated with the MMIC. These initial results are vindicated by more detailed analyses
	investigating the solar activity dependence of MM location and intensity 
	following the $B$-field-only study of \citeA{volwerk_mirror_2016}.
	
	At Mars, \citeA{bertucci_mgs_2004} proposed that certain low-frequency wave signatures in the NASA/Mars Global Surveyor (MGS) dataset, based on an anticorrelation between electron fluxes and magnetic field intensities, could be consistent with MM behaviour. However, this could not be conclusively demonstrated as MGS did not include any ion analyser and thus the MMIC could not be checked. The same year, \citeA{espley_observations_2004} studied the magnetic fluctuations at Mars using MGS magnetometer data to argue for the presence of MM structures in the dayside magnetosheath. Because the ESA/Mars Express spacecraft did not include any magnetometer, no MM study could be performed in the interim leading to the NASA/Mars Atmosphere and Volatile EvolutioN (MAVEN) mission launched in 2014. Recently, using time-frequency analysis of magnetometer and plasma data from MAVEN, \citeA{ruhunusiri_low-frequency_2015} created maps of the low-frequency wave content of the Mars plasma environment for the first year of operations, using the hierarchical identification scheme of \citeA{song_identification_1994} based on transport ratios. Whereas slow mode waves (Alfvén and quasi-parallel slow modes) were statistically found to dominate both in the solar wind and in the magnetosheath region, waves consistent with mirror modes were on average confined on the dayside to the region closest to the magnetic pile-up boundary (MPB) and extending on the nightside in the magnetotail. This would seem roughly consistent with one theoretical picture that MM structures, if originating upstream in the solar wind as so-called {\it magnetic holes} (MH) where they are routinely detected \cite{madanian_magnetic_2020}, need time to grow when crossing the bow shock (BS) region \cite{ahmadi_simulation_2017}. They may then become more prominent downstream of the shock when they are convected away by the shocked solar wind plasma. Since, in \citeA{ruhunusiri_low-frequency_2015}, both quasi-perpendicular and quasi-parallel bow shock crossings were mixed, conditions for MM wave generation were difficult to assess and will need further study. Different scales of MHs are also present in the magnetosheath of Mars as recently reported by \citeA{wu_statistical_2021}, with the near-ubiquitous presence of small-scale linear MHs in the magnetosheath of Mars lasting less than \qty{1}{s} in the data. 

	The MAVEN mission, including dedicated magnetometer \cite{connerney_maven_2015} and high-time resolution plasma instruments \cite{halekas_maven_2015,mcfadden_maven_2015}, provides a unique opportunity to study these structures in more detail at a planet where, comparatively to other magnetised and non-magnetised objects only little study has been performed as of now.
	Because (i) plasma bulk speeds in the magnetosheath at Mars and Venus are of similar magnitude reaching a large fraction of the upstream solar wind speed \cite{halekas_flows_2017,bader_proton_2019}, 
	and (ii) MAVEN has a relatively slower orbital speed of \qty{0.6\times10^{-3}}{$R_P$/s} with $R_p$ the planet's radius, i.e., bow shock crossings at about \qty{2}{km/s} compared to VEX' \qty{1.5\times10^{-3}}{$R_P$/s}, MM structures are expected to appear in the MAVEN data on at least similar timescales as for VEX' detections.
	
	To our knowledge, we present in this study the first complete observation of quasi-linear MM structures at Mars. This is made possible, for the first time, with the high-temporal resolution enabled by MAVEN's ion and electron measurements (\qty{4-8}{s} and \qty{2}{s} respectively), in contrast to all previous missions, which had a too limited payload and/or temporal resolution. Our event took place during the early part of the MAVEN mission on 25 December 2014 around 11:30\,UT, in the receding phase of solar cycle 24. MMs were found wedged against the MPB behind a quasi-perpendicular bow shock crossing --- a classical empirical picture reminiscent of Earth-bound observations. After describing our detection methods based on magnetometer data only at \qty{1}{-Hz} sampling frequency and validated against calculated ion and electron plasma moments (Section\ \ref{sec:methods}), we present an in-depth analysis of these MM structures in Section\ \ref{sec:inDepthAnalysis}, providing recommendations for a new set of $B$-field-only detection criteria (Section\ \ref{sec:reanalysis}). Finally, we discuss their possible origin in Section\ \ref{sec:discussion}. One of the purposes of this study is to prepare for a statistical analysis of MM occurrences using the whole MAVEN dataset.

\section{Data analysis}\label{sec:methods}

	\subsection{Instrumentation}\label{sec:instrumentation}
	
	The MAG package consists of two tri-axial fluxgate magnetometers mounted at the extremity of two boomlets on MAVEN's solar panels \cite{connerney_maven_2015}. It measures $3$-component magnetic fields with a nominal frequency of \qty{32}{Hz} and an accuracy better than $0.05\%$. Such high temporal resolution is not necessary to investigate the global behaviour of MM structures and hence a $1$-s resolution is adopted throughout.

	As part of the Particles and Fields package on board MAVEN, the Solar Wind Ion Analyzer (SWIA) is an electrostatic ion analyser measuring ion differential fluxes with a maximum temporal resolution of \qty{4}{s} \cite{halekas_solar_2015}. Sensitivity is automatically switched to accommodate for a large dynamic range of measured ion fluxes (magnetosphere and solar wind). SWIA has thus two three-dimensional (3D) scanning modes, one coarse ($360\deg\times90\deg$, $\Delta\phi=\Delta\theta=22.5\deg$ in anode/deflection angles, energy \qty{25}{eV/q} to \qty{25}{keV/q} \cite{halekas_structure_2017}, suitable for magnetosphere and pickup ions), one fine ($45\deg\times45\deg$, $\Delta\phi=4.5\deg$, $\Delta\theta=3.75\deg$, $10\%$ energy windows, suited to solar wind ions), both with $48$ energy steps. 
	Additionally, two separate 3D data packet telemetry modes are used on board which result in the Survey (lower cadence, maximum availability) and Archive (higher cadence but lower availability) modes. The combination of the scanning and telemetry modes results in four different data products, labelled in the following SWIFA (SWIA Fine Archive), SWIFS (Fine Survey), SWICA (Coarse Archive) and SWICS (Coarse Survey). The last two modes are thus those with lowest angular resolution and are usually selected in the magnetosheath. 
	
    Each of these four data products yields different moments at different temporal resolution. Onboard moments are automatically calculated either in coarse or fine modes, making use of the highest cadence available: in practice, they are usually provided at the highest possible temporal resolution of \qty{4}{s}. Because SWIA does not discriminate between ion masses, the plasma composition needs to be assumed first \cite{halekas_solar_2015}. Discrimination in the solar wind between H$^+$ and He$^{2+}$ ions can be done on an energy-per-charge basis. Depending on composition assumptions, errors of the order of $\sqrt{M}$ are introduced on the onboard-calculated moments, $M$ being the mass of the plasma (for example, $M=$\qty{1}{amu} for protons) \cite{halekas_structure_2017}. The proton-dominated composition assumption is usually fulfilled in the not-so-dense parts of the Martian plasma environment, i.e., the solar wind and magnetosheath down to the outer edge of the Magnetic Pile-up Boundary (MPB, sometimes also referred to as Induced Magnetosphere Boundary or IMB), which the present study concentrates on. In regions where heavy ions dominate (as in the ionosphere and during deep-dip campaigns), it is however necessary to call in supporting instruments with mass-resolving capabilities, such as the Suprathermal and Thermal Ion Composition analyser (STATIC) \cite{mcfadden_maven_2015}. In the following, ion moments of order 0 (density $N_i$ in cm$^{-3}$), 1 (velocity vector $\v V$ in km\,s$^{-1}$) and 2 (pressure tensor $\overline{\overline{\v P}}$ in eV\,cm$^{-3}$ and associated temperatures $T$ in eV) are used. Calculations of these moments depend also on the energy thresholds and extreme care must be achieved in their derivation \cite{halekas_structure_2017}.

    The Solar Wind Electron Analyzer (SWEA), located on a 1.5-m long boom, is a hemispheric electrostatic electron analyser designed to measure the energy and angular distribution of photo- and solar-wind electrons in the $3$--\qty{4600}{eV} energy range in $64$ bins, with a resolution of $\Delta E/E \sim 17\%$ and a maximum cadence of \qty{2}{s} \cite{mitchell_maven_2016}. It swipes almost all \qty{4\pi}{sr} of solid angle with a $22.5\deg$ angular resolution in the azimuth direction and $20\deg$ along the direction of the elevation angle. As MAVEN is a 3-axis stabilised spacecraft, the field of view (FOV) is broadened up to $360\deg\times120\deg$ by the use of electrostatic deflectors theoretically covering $87\%$ of the sky, but reduced to $79\%$ due to spacecraft obstruction. 
    Omnidirectional energy spectra (product `SPEC') are calculated onboard at the highest temporal resolution by integrating the 3D distribution over angular sectors. Densities presented in the following are calculated from these onboard energy fluxes by integrating over the $64$ energy bins, assuming isotropy over SWEA's blind spots.
    With its temporal resolution of \qty{\sim2}{s}, SWEA complements SWIA for total plasma density measurements when structures lasting less than about \qty{8}{s} ($2\times\Delta t_{\rm SWIA}$) are identified in magnetometer datasets, which is often the case for MMs. Because of quasi-neutrality, the electron density $N_e$ must theoretically be equal to the ion density, i.e., $N_e\sim N_i$. The absolute determination of the electron density depends on different experimental parameters (such as the product used) and on the accuracy in the determination of the spacecraft potential. The electron density from the SPEC product has the better available time resolution but its absolute value can differ from the moments computed on the ground from the 3D velocity distribution. Hence, only absolute variations in electron densities may be used in those cases to corroborate ion density variations. 
	
    With the help of MAG, SWEA and SWIA, several temporal and spatial scales are probed, incrementally accessing plasma features at \qty{1}{s} and below, to \qty{2}{s}, to \qty{4\text{--}8}{s}, respectively.
    
	The coordinate system adopted in the following for all vectorial and tensorial quantities such as $\v B$, $\overline{\overline{\v P}}$ or $\v V$, is that of Mars Solar Orbital coordinates (also known as Sun-state coordinates), abbreviated MSO: centred on Mars, the $X$ axis points towards the Sun from the planet's centre, $Z$ is in the direction of the North pole from the orbital plane, and $Y$ completes the right-hand triad, so that the $X$--$Y$ plane is Mars' orbital plane around the Sun. All quantities are thus expressed in MSO coordinates.

    \subsection{Detection criteria: $B$-field only}\label{sec:CriteriaB-fieldOnly}
	
	Because of the ubiquity of magnetometers on board space missions (at the notable exception of Mars Express) and their usually high sampling frequency, mirror mode structures have traditionally been located by $B$-field only measurements \cite{lucek_mirror_1999,joy_mirror_2006,soucek_properties_2008,volwerk_mirror_2016}, relying on compressional and linearly polarised signatures. However, structures such as foreshock waves or fast mode waves linked to pickup ions are compressional in nature and may also fulfill these criteria. In those cases, plasma measurements are necessary to lift the ambiguity; one way is to check for the anticorrelation between density and magnetic field variations that is expected for MM structures \cite{hasegawa_drift_1969}. Another is to evaluate, when possible, the MM-unstable condition with the help of the MMIC, assuming that the spacecraft crosses the source region of the MMs. In the following, $B$-field-only criteria are first used to pinpoint promising structures in the magnetosheath. They are then validated against plasma measurements (Section\ \ref{sec:criteriaValidation}).    
	
	Table\ \ref{tab:Bcriteria} lists the MM detection criteria chosen, with references that inspired them. Following previous studies at Venus,  only quasi-linearly polarised waves are sought. As in \citeA{volwerk_mirror_2016}, magnetic field measurements are first low-pass filtered using a $2$-min-wide Butterworth filter ($f_{\rm band} = 1/120$\,Hz, passband ripple $2$\,dB and stopband attenuation $20$\,dB) to approximate the total background field $\left|\v B_\text{bg}\right| = B_\text{bg}$. A compressional signature consistent with MMs is defined by large fluctuations around the background field, i.e., $\Delta |{\v B}|/B_{\rm bg} = \left||\v B|-B_\text{bg}\right|/B_\text{bg}$, with threshold values typically ranging from $0.1$ to $0.2$. For example, \citeA{volwerk_mirror_2016} used a threshold of $\Delta |{\v B}|/B_{\rm bg} \geq 0.20$.	
	When dip and peak mirror structures are both present in the studied interval, \citeA{joy_mirror_2006} advocated the use of $B$-field lower and upper quartiles to determine the `true' background level: this may be an issue especially when long portions of sinusoidal-like oscillations lead to an under- or over-estimation of the ambient field intensity and hence a correspondingly smaller $\Delta |{\v B}|/B_{\rm bg}$. We have checked that this issue does not significantly affect the results presented here by lowering the absolute fluctuation $\Delta |{\v B}|/B_{\rm bg}$ threshold from $0.20$ to $0.15$ in order to potentially capture more events. 

	A minimum variance analysis \cite{sonnerup_minimum_1998} (MVA) on the magnetic field directions is then performed with a $15$-s moving window. The aim of MVA for MM detection is to constrain the wave polarisation of the detected structures \cite{lucek_mirror_1999}. Ideally, linear structures are sought, adopting a cigar-shaped variance ellipsoid with maximum and intermediate eigenvalues $\lambda_\text{max}$ and $\lambda_\text{int}$ clearly distinct from one another \cite{genot_kinetic_2001,tatrallyay_evolution_2002}. In practice, non-linear structures have also been observed and studied theoretically \cite{califano_nonlinear_2008}, with elliptically polarised MMs found more often than stricto sensu linearly polarised ones \cite{genot_kinetic_2001,genot_mirror_2008}.
	Consequently, criteria based on eigenvalue ratios  
	should not be too constraining if one wants to capture candidate events straying somewhat from the pure linear wave polarisation \cite{sergeev_survey_2006}, hence the ``quasi-linear'' nomenclature adopted here. In parallel, the quasi-linear nature of the sought-after waves can be reinforced by calculating the angle $\Theta_\text{maxV}$ ($\Phi_\text{minV}$) between the background field direction and the maximum (minimum) variance directions (i.e., eigenvectors). Threshold values on these angles were adopted from \citeA{volwerk_mirror-mode-like_2008} and \citeA{volwerk_mirror_2016}.
	
	Quasi-linear MM wave candidates are finally automatically detected from $B$-field measurements only if the criteria listed in Table\ \ref{tab:Bcriteria} are fulfilled: (i) compressional structures (criterion $1$), (ii) quasi-linear wave polarisation (criteria $2-4$) and (iii) sufficiently large average $B$-field intensities in a $\pm$\qty{2}{min} interval around the MM candidate (magnetosheath conditions, criterion $5$). Values for each criteria have been determined empirically starting from theoretical considerations \cite{price_numerical_1986}, past observations at Earth and Venus, and tweaked to the MAVEN Mars dataset to capture at least once most structures seen in a test interval. Criterion $5$ ($\langle B_{\rm bg}\rangle\geq 11$\,nT), inspired by a similar one from \citeA{dimmock_statistical_2015} for observations at Earth,
	aims at removing potential outliers outside of typical magnetosheath $B$-field levels in an interval around the MM candidate. 
	
	Non-Gaussian statistics may additionally help with the classification of MM candidates \cite{osmane_universal_2015}. The propensity of the signal distribution in a given time interval (containing, for instance, $m$ successive measurements of $B$) towards peaks or dips can be measured by its skewness, expressed as ${\rm Skew} = \frac{1}{m} \sum_{j=1}^m{\left(|{\v B_j}|-B_{\rm bg}\right)^3}/\sigma^3$, where $\sigma$ is the standard deviation of the module $|{\v B_j}|$ of ${\v B}$ at each time step $j$. 
	A positive skewness denotes the presence of peaks whereas a negative skewness points to the presence of dips (criterion $6$ in Table\ \ref{tab:Bcriteria}). Similarly, the excess kurtosis, defined as ${\rm eKurt} = \frac{1}{m} \sum_{j=1}^m{\left(|{\v B_j}|-B_{\rm bg}\right)^4}/\sigma^4 - 3$, gives a measure of the fluctuations of the signal (presence of outliers in the tail of the probability distribution function): a negative kurtosis, significantly different from $0$ (pure Gaussian statistics), implies a $B$-field signal with large oscillations, a characteristic that trains of MMs share \cite{osmane_universal_2015}. This is given in Criterion 6.
	
	Finally, magnetic azimuth and elevation angles, defined as $az = \arctan{\left(B_y/B_x\right)}$ and $el = \arctan{\left(B_z/\sqrt{B_x^2+B_y^2}\right)}$, can be calculated. The magnetic field is expected to rotate by less than about $15\deg$ for linear MMs \cite{treumann_strange_2004}, a property that can be checked after the detection is performed.
	
    \begin{table}
        \caption{Magnetic field-only criteria for detection of mirror modes and their classification. The absolute $B$-field fluctuation is estimated by $\Delta |{\v B}|/B_{\rm bg}$, with $B_\text{bg}$ the background intensity (low-pass Butterworth filter). A moving MVA analysis with a $15$-s window is performed. $\Theta_\text{maxV}$ ($\Phi_\text{minV}$) is the angle between the maximum (minimum) variance direction and that of the background field. $\lambda_\text{max}/\lambda_\text{int}$ is the ratio of the maximum and intermediate eigenvalues, whereas $\lambda_\text{int}/\lambda_\text{min}$ is the ratio of the intermediate and minimum eigenvalues, as determined from the MVA analysis. Criterion 5, with average $B$-field $\langle B_{\rm bg}\rangle$ performed on an interval $\pm$\qty{2}{min} around the potential candidate MM, further checks for potential outliers that are tentatively in the solar wind. Criterion 6 classifies MMs into predominantly dips and peaks in the time interval considered (for example $\pm$\qty{2}{min} around the detected candidate). Criterion 7 checks for changes of magnetic azimuth and elevation angles across an individual MM structure $\Delta (az,el)$. The {\it initial values} are those used for the first detection of the structures (Section\ \ref{sec:inDepthAnalysis}), whereas the {\it revised values} were manually obtained after validation of the detections (Section\ \ref{sec:reanalysis}).
        }
        \centering
        \footnotesize
        \begin{tabular}{c l c c l l}
            \hline
            $\#$  & Criterion 		            & Initial Value  & 	Revised		& Reason 					& Example reference(s)  \\
            \hline
            $1$  & $\Delta |{\v B}|/B_{\rm bg}$ & $\ge 0.15$ 	& $\ge 0.14$	& Compressional structure	&  \citeA{genot_mirror_2009, volwerk_mirror_2016} \\
            $2$  & $\Theta_\text{maxV}$         & $\le 20^\circ$  & $\le 23^\circ$ & Linearly polarised waves 	& \citeA{lucek_identification_1999,volwerk_mirror_2016}\\
                 & $\Phi_\text{minV}$   	    & $\ge 70^\circ$	& $\ge 70^\circ$ & Linearly polarised waves 	& \citeA{lucek_identification_1999,volwerk_mirror_2016}\\
            $3$  & $\lambda_\text{max}/\lambda_\text{int}$  & $\geq3$ &  $\geq2.5$	& Linearly polarised waves & \citeA{genot_kinetic_2001,soucek_properties_2008}\\
            $4$  & $\lambda_\text{int}/\lambda_\text{min}$  & $\leq6$ & $\leq8$ & Linearly polarised waves & 
   \citeA{genot_kinetic_2001,soucek_properties_2008}\\
            $5$  & $\langle B_{\rm bg}\rangle$  & $\ge\qty{11}{nT}$	& $-$ & Magnetosheath conditions	&  \citeA{dimmock_statistical_2015} \\
            \hline
            $6$  & $\rm Skew$   & $< 0$           &  & dips         & \citeA{osmane_universal_2015,dimmock_statistical_2015}\\
                 & 	            & $> 0$                   &. & peaks      & \citeA{osmane_universal_2015,dimmock_statistical_2015}\\
                 & $\rm eKurt$ & $\leq 0$            & & non-Gaussian fluctuations &   \citeA{osmane_universal_2015}\\
            $7$  & $\Delta(az, el)$ & $\leq 15^\circ$   & & Linearly polarised waves & \citeA{treumann_strange_2004,tsurutani_magnetosheath_2011} \\
            \hline
        \end{tabular}
        \label{tab:Bcriteria}
    \end{table}

    Note that the eigenvalue criteria (Criteria 3 and 4) aim at reinforcing the quasi-linear polarised nature of the wave modes, ensuring that the maximum variance direction (the tangential component of the eigenvector triad) is well defined and that quasi-degeneracy of the covariance matrix is kept to the two minimum eigenvalues. Although the direction of the wave vector $\v{k}$ is in general ill-determined, criteria $3-4$ imply that the variance ellipsoid is in general cigar-shaped, and elliptically-polarised structures are mostly filtered out to the benefit of linearly polarised waves.

    \subsection{Validation with plasma measurements}\label{sec:criteriaValidation}

	In parallel, SWIA plasma densities $N_i$ for the available modes are calculated from the moments of order $0$ of the ion distribution function. Moreover, the background-normalised amplitude $\Delta N/N = (N_i - N_\text{bg})/N_\text{bg}$ is calculated where $N_\text{bg}$, the background plasma density, is obtained by applying a low-pass Butterworth filter to the SWIA $N_i$ densities, with the same filter parameters as for the magnetic field.  In this way, $\Delta N/N$ has the advantage of changing signs for peaks or dips in the signal. For comparison, total $B$-field intensities are downsampled to SWIA's mode resolution and normalised amplitudes are calculated in the same way, i.e.,  $\Delta B/B_\text{bg} = \left(|{\v B}| - B_\text{bg}\right)/B_\text{bg}$. When $\Delta B/B$ and $\Delta N/N$ are of opposite signs at a candidate MM, they are in effect anticorrelated. Those candidates that fulfill the $B$-field-only criteria and the anticorrelation signature are considered MM signatures proper. 

    For further confirmation of the nature of the MM candidates, two separate criteria are further checked against: 
    \begin{itemize}
        \item Direction (also, Criterion $7$ in Table\ \ref{tab:Bcriteria}): the magnetic field direction in a region containing MMs should not change by more than $15^\circ$ on average. This stems from theoretical considerations \cite<due to a small parallel wave vector component, see>{treumann_strange_2004,tsurutani_mirror_2011} and the fact that MM, as linearly polarised waves, are predominantly growing at large angles to $\v{B}$. Such directional change can be measured by the azimuth and elevation angles of the magnetic field, as shown in Figure\ \ref{fig:MMdetection}.
        \item Instability criterion MMIC (Equation\,\ref{eq:instabilityCriterion}): because large uncertainties in deriving reliable temperatures from the diagonalisation of the higher-order moment (the pressure tensor) may arise due to composition assumptions, caution must be exercised when calculating MMIC in the inner magnetosheath and, especially, close to the MPB, where heavier ions may be present. In the time span where MM-like structures were found in the present study (11:26--11:30\,UT), STATIC energy-mass spectrograms agreed with SWIA and additionally showed that both He$^{2+}$ and heavier ions such as O$^+$ were present in the higher energy-per-charge levels. However, choosing a reduced energy threshold to restrain the contamination did not significantly impact the determination of the moments of the ion distribution function with SWIA. Consequently, in this study, the moments were calculated in the full range \qty{25-25\times10^3}{eV/q}.
    \end{itemize}
    
    In order to calculate the MMIC criterion from Equation\,(\ref{eq:instabilityCriterion}), ion pressures $P_{||}$ and $P_\perp$ in the parallel and perpendicular direction with respect to the ambient $\v B$ -field must be first estimated. This is done using the full pressure tensor calculated as the moment of order $2$ of the ion distribution function from the appropriate SWICA mode (maximum angular resolution available). Temporal resolution of the SWICA mode for this study is \qty{8}{s}. The pressure tensor is symmetric by construction and consists of $9$ elements, i.e., $3$ diagonal elements $P_{xx}$, $P_{yy}$ and $P_{zz}$, and a total of $6$ off-diagonal elements $P_{xy} = P_{yx}$, $P_{xz} = P_{zx}$ and $P_{yz} = P_{zy}$.  
    
    Two methods are usually adopted to obtain $P_{||}$ and $P_\perp$. They are presented in more detail in \ref{appendix2}. In short, Method $1$ uses a direct diagonalisation of the pressure tensor, whereas Method $2$ first rotates the pressure tensor into a Mean Field-Aligned coordinate system (MFA, see \ref{appendix1}) with one direction along the mean $\v B$-field and diagonalises the remaining $2\times2$ tensor to obtain the two perpendicular directions \cite{halekas_flows_2017}. Method $1$ may provide a quick first estimate of directions but implicitly assumes that one of the principal axes of the tensor ellipsoid is aligned with the magnetic field direction, an hypothesis that is usually false. Moreover, off-diagonal terms still present in the MFA coordinate system have physical significance: indicative of non-gyrotropy \cite{swisdak_quantifying_2016,che_quantifying_2018}, they correspond to shear stresses in the ambient plasma, which Method $1$ ignores. Thus, Method $2$ is preferred in the following. The MFA direction is evaluated by taking the low-pass filtered $\v B_\text{bg}$ (as calculated in Section\ \ref{sec:CriteriaB-fieldOnly}) since the average field over the SWICA measuring time of \qty{8}{s} (taken as \qty{\pm4}{s} around the middle of the scanning interval) gives almost identical results in the time span considered in this study (\ref{appendix1}).
    
    As discussed in \citeA{halekas_flows_2017}, the limited field of view (FOV) of the SWIA instrument ($360\deg\times90\deg$ for the SWICA mode) impacts the quality of the retrieval. At times when the ambient magnetic field direction is in the FOV of SWIA, one of the perpendicular components may be difficult to estimate (the maximum perpendicular eigenvalue of the pressure tensor taken as an upper limit). At times when it is not in the instrument's FOV, the derivation of the parallel pressure, and hence parallel temperature, will suffer from comparatively larger uncertainties. In the interval of study where MM-like structures were detected (2014-12-25 11:26--11:30\,UT,), the background magnetic field direction was unfortunately not in the FOV of the instrument (with angles between the $Z$ axis of SWIA and the magnetic field direction exceeding $135\deg$, see Figure\ \ref{fig:MMevent1_contextGeneral}). This made the determination of the parallel component of the pressure tensor difficult. As a result $P_{||}$ was likely underestimated, as the flux integration over all angles could not capture the full 3D ion velocity distribution function (VDF) in that direction. However, magnetosheath ion VDFs are not beam-like and may spread over several angular sectors of SWIA. A cursory examination of the ion VDFs during this interval showed that the peak of the 3D distribution (in the bulk plasma direction) was within SWIA's FOV, so that the retrieved parallel component of the pressure tensor was more representative of the core contribution of the ions. As such, they represent a good proxy of $P_{||}$ and $T_{||}$.
    
    The final derivation from the pressure tensor of the parallel and perpendicular temperatures ($T_{||} = P_{||}/N_i$ and $T_\perp = P_\perp/N_i$ with the temperatures in eV), as well as the plasma-$\beta_\perp$ with respect to the background field $\v B_\text{bg}$ assumes only one ion species (protons), enabling the estimation of the MMIC.    
    For a faster and automatic detection bypassing the MMIC criterion, the normalised (electron and ion) density variations in anticorrelation with $|B|$-field variations proved to be an adequate and sufficient tool to confirm the presence of MM structures during the time intervals considered in this study.

\section{Observations}\label{sec:Observations}

	\subsection{General description}\label{sec:contextObservations}
	
	\begin{figure}
		\noindent\includegraphics[width=0.65\textwidth]{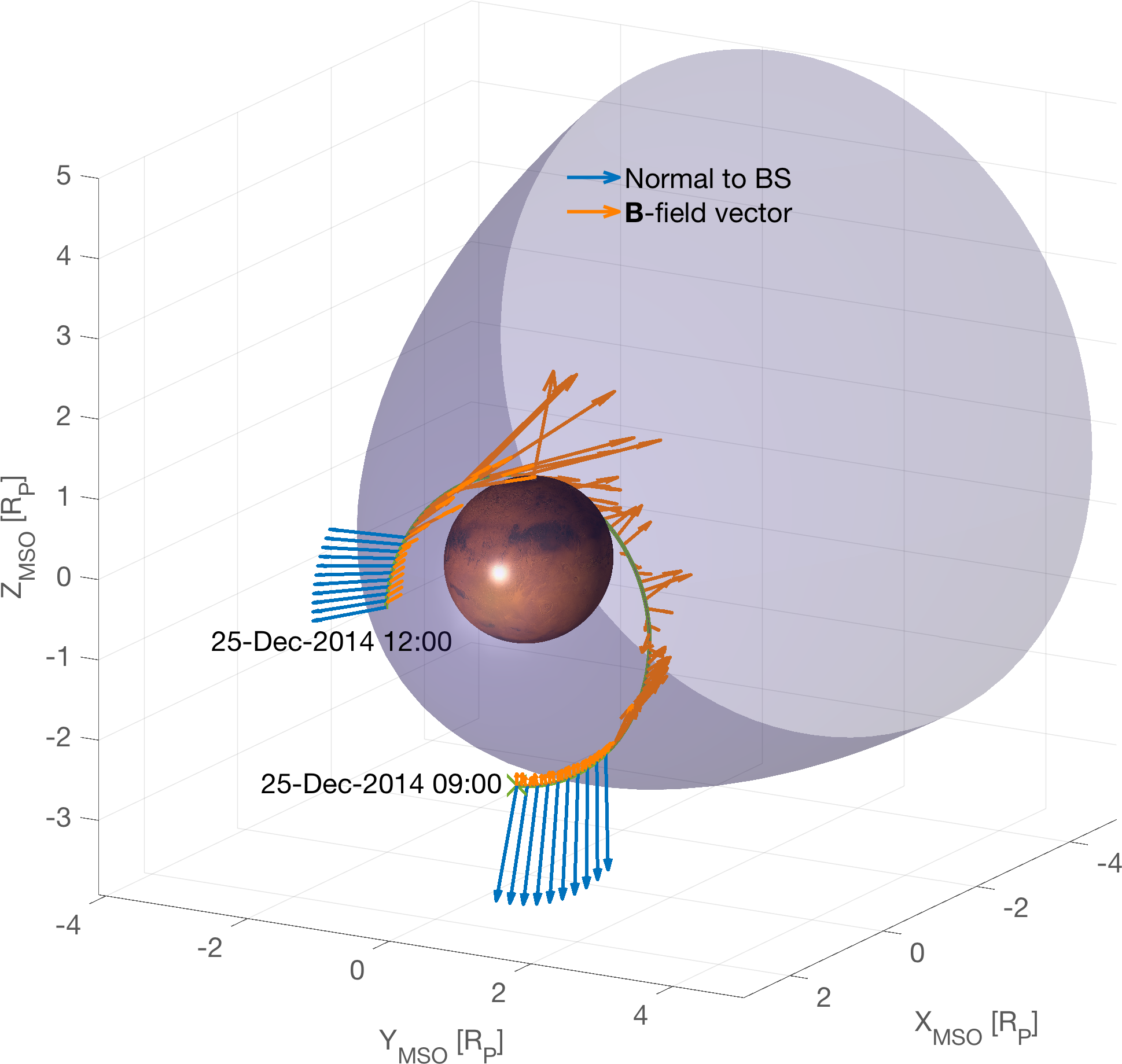}
		\caption{Event configuration with MAVEN orbit in blue from 09:00 to 12:00\,UT on 25 December 2014, in MSO coordinates (units normalised to Mars' mean radius $R_p$). The $X_\text{MSO}$ axis points towards the Sun. The vectors normal to the bow shock are drawn as blue arrows, whereas the average magnetic field directions during the orbit are drawn as orange arrows. The bow shock model shown and used for the bow shock normal calculations is that of \citeA{gruesbeck_three-dimensional_2018} (all points).
		}
		\label{fig:MMevent1_general}
	\end{figure}
	
	On 25 December 2014 around 09:46\,UT and around 11:40\,UT, two prominently quasi-perpendicular bow shock crossings took place. The first of these crossings was extensively studied by \citeA{Burne2021} who noted the quiet upstream solar wind conditions with no coronal mass ejection (CME), or other solar wind transients. These authors concluded that the mass-loaded shock around those times exhibited well-defined supercritical features, such as foot, ramp and overshoot, with scales reminiscent of those found at Earth, making it a baseline example at Mars. The magnetosheath behind these two quasi-perpendicular shocks is thus expected to be in a so-called `quiet' state, with no external disturbance in the solar wind flow. 	
	
	The corresponding orbit of MAVEN during those two crossings is shown in MSO coordinates in Figure\ \ref{fig:MMevent1_general} between 09:00 and 12:00\,UT, and shown together with the statistical position of the bow shock (BS), the direction of the background magnetic field (green) and that of the normal to the shock surface (blue). The normal to the BS was calculated using the analytical model of \citeA{gruesbeck_three-dimensional_2018} 
	assuming the shock surface to be smooth. Moving first from the solar wind to the magnetosheath, MAVEN resided approximately $1$ hour in the magnetosphere, circling the planet and exiting again into the upstream solar wind around 11:40:16\,UT in the $(+Y_{\rm MSO},+Z_{\rm MSO})$ dayside quadrant.
    The angle between the average magnetic field and the normal to the bow shock was found for this crossing to be $\theta_\text{Bn} = 76^\circ$, assuming a smooth bow shock surface.

\begin{figure}
		\noindent\includegraphics[width=\textwidth]{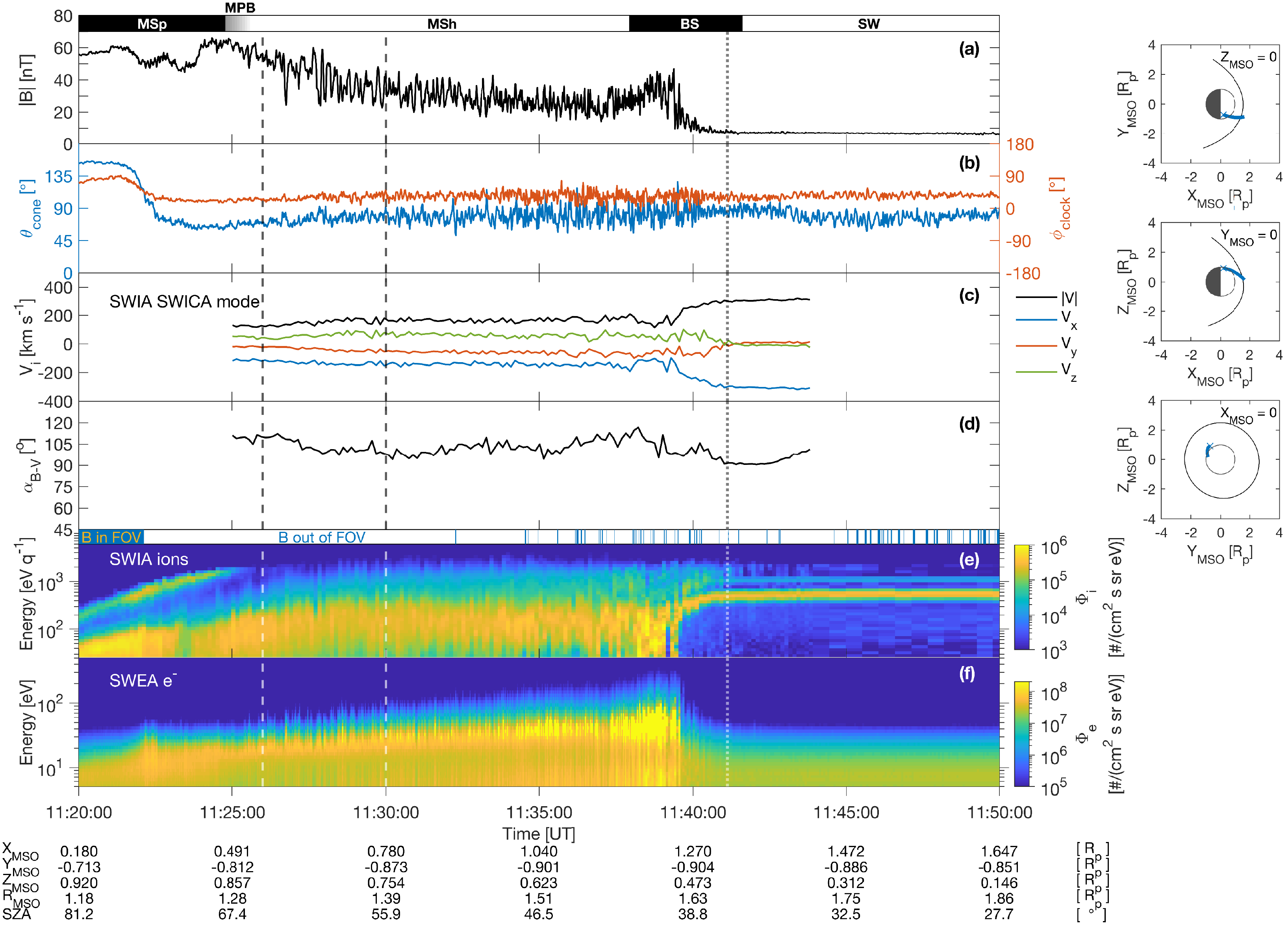}
		\caption{Magnetic field, ion and electron plasma measurements measured with MAVEN, 25 December 2014 between 11:20 and 11:50\,UT. (a) Module of the magnetic field from MAG, (b) magnetic clock and cone angles, (c) ion velocity vector measured by SWIA (SWICA mode), (d) angle between the mean magnetic field $\v B_\text{bg}$ and the ion velocity $\v V_i$ directions $\alpha_{B-V}$ at SWIA's 8-s resolution, (e) ion flux spectra measured by SWIA (SWICA mode), (f) electron flux spectra measured by SWEA. Energy-time spectra show the particle differential fluxes $\Phi_i$ and $\Phi_e$ in cm$^{-2}$\,s$^{-1}$\,sr$^{-1}$. Identified regions in Mars' plasma environment are as follows: solar wind (SW), bow shock, its substructures and wake (BS), magnetosheath (MSh) and magnetosphere (MSp). The position of the bow shock is automatically estimated via a simple geometrical algorithm based on the model of \citeA{gruesbeck_three-dimensional_2018} 
		and shown as a vertical dotted line around 11:41\,UT. The two vertical dashed lines encompass the interval where clear mirror mode signatures were found, as explained in the text. SWICA velocities are valid mostly inside the magnetosheath, hence no value is shown inside the MPB and far upstream of the shock. Immediately above the SWIA spectra, the associated field-of-view (FOV) information is given with blue regions showing when the $B$-field direction is in the FOV of the instrument, and white regions when it is not. 
		Right:  Sections in $X_{\rm MSO}-Y_{\rm MSO}$, $X_{\rm MSO}-Z_{\rm MSO}$ and $Y_{\rm MSO}-Z_{\rm MSO}$ planes of MAVEN's orbit during that time, with the cross representing the starting point at 11:20\,UT. The bow shock shape model in black is that of \citeA{gruesbeck_three-dimensional_2018} (all points) and is represented on the panels on the left by a dotted vertical line. Coordinates ($X_{\rm MSO}, Y_{\rm MSO}, Z_{\rm MSO}$, planetocentric distance $R_{\rm MSO}$), expressed in units of Mars' planetary radius $R_p$,  as well as the solar zenith angle (SZA) are indicated for a few time stamps. 
		}
		\label{fig:MMevent1_contextGeneral}
	\end{figure}

    Figure\ \ref{fig:MMevent1_contextGeneral} presents an overview of magnetic field and plasma measurements between 11:20 and 11:50\,UT. Panels (a) and (b) show the total magnetic field intensity $|B|$ and magnetic cone and clock angles as a measure of the field's direction, defined as $\theta_{\rm cone} = \arctan{\left(\sqrt{B_y^2+B_z^2}/B_x\right)}$ and $\phi_{\rm clock} = \arctan{\left(B_z/B_y\right)}$. A cone angle of $0\deg$ ($180\deg$) implies a sunward (antisunward) $B$-field direction, whereas a clock angle of $0\deg$ ($90\deg$) indicates a direction in the $+Y_{\rm MSO}$ ($+Z_{\rm MSO}$ ) direction. Several regions encountered by MAVEN in its exit towards the upstream solar wind are highlighted: after spending some time in the magnetosphere (labelled `MSp') with large rotations of the field first directed roughly antisunward and then abruptly rotating towards the sunward direction, magnetic field fluctuations become attenuated indicating the presence of the MPB (depicted as a grey zone) when transiting towards the magnetosheath (labelled `MSh') around 11:24:45\,UT. In the magnetosheath, several quasi-periodic dips and peaks are observed in sequence in the magnetic field data, about ten minutes before the quasi-perpendicular BS crossing into the solar wind (`SW'). Field directions remain on average fairly constant in the magnetosheath, but tend to fluctuate more and more when the spacecraft closes in on the shock structure. In the upstream solar wind, no indication of MHs was found.

    Panel (c) of Figure\ \ref{fig:MMevent1_contextGeneral} shows for reference the ion velocity measured by SWIA in SWICA mode, which is valid mostly in the magnetosheath. The plasma is moving in the magnetosheath at velocities of about \qty{150}{km/s} on average, with little change in direction and amplitude until the spacecraft exits into the upstream solar wind around 11:40\,UT. Panel (d) calculates the angle, noted $\alpha_{B-V}$, between the background magnetic field direction (downsampled to SWIA's 8-s resolution) and the ion velocity vector measured by SWIA. The background magnetic field $\v B_\text{bg}$ is obtained from a low-pass Butterworth filter of the 1-s $B$-field data, as explained in Section\ \ref{sec:CriteriaB-fieldOnly}. The angle remained close to about $100\deg$ with little variation around that value ($\pm 6\deg$) during the entire interval shown here, that is, the plasma flow and the magnetic field stayed quite perpendicular to each other throughout. This could result in ion pickup ring distributions favouring the emergence of instabilities in the solar wind as well as downstream of the shock \cite{Price1989}.

    Panels (e) and (f) display the ion and electron differential flux spectra from SWIA and SWEA (omnidirectional), respectively, with $\Phi_j$ the differential particle flux of charged species $j$ (with $j$=[ions,e$^-$]). It is important to remark that SWIA's field of view during this interval does not encompass the direction of the magnetic field, resulting in a likely underestimate of the calculated moments along the parallel direction to the field (See Section\ \ref{sec:inDepthAnalysis}). In the magnetosphere ('MSp'), two separate ion populations of different energies but with rather constant fluxes can be seen, one below \qty{100}{eV/q}, the other above \qty{400}{eV/q}, likely mixing fast protons and heavier species produced by EUV photoionisation of Mars' ionosphere. This coincides with lower electron fluxes with energies around \qty{20-25}{eV}, typical of photoelectron energies. In the magnetosheath ('MSh'), the ion spectra broaden in energy with a main population of protons at a mean energy of about \qty{200}{eV/q} and large flux fluctuations that appear related to the large magnetic field fluctuations seen then. Similar effects are seen on the electron spectra, with an intensification of electron energies and fluxes together with a high flux variability. Ion differential fluxes reach about \qty{4\times10^{5}}{cm$^{-2}$\,s$^{-1}$\,sr$^{-1}$\,eV$^{-1}$} whereas mean electron fluxes are around \qty{10^{8}}{cm$^{-2}$\,s$^{-1}$\,sr$^{-1}$\,eV$^{-1}$}.

    Closer to the bow shock around 11:37\,UT, in the turbulent wake and transiting into the solar wind, the magnetic field varies a lot and a steady drop in ion energies with an increase in ion fluxes can be seen. This ion energy drop is concomitant with a broadening of the electron energy distribution, with electron fluxes above \qty{2\times10^{8}}{cm$^{-2}$\,s$^{-1}$\,sr$^{-1}$\,eV$^{-1}$} around \qty{25}{eV} energy. In the solar wind, two stable main populations of ions are present, interpreted as protons (yellow-orange broad line) and He$^{2+}$ particles (blue-green thinner line) at about $600$ and \qty{1200}{eV/q}, equivalent to a bulk speed of \qty{\sim340}{km/s}.
    
    The upstream solar wind conditions after 11:41 UT were nominal in the sense of \citeA{slavin_solar_1981}, with \qty{|B|\approx 6.6}{nT}, \qty{V_\text{sw}\approx 340}{km/s}, \qty{T_i\approx 7}{eV}, \qty{T_e\approx12}{eV} and \qty{n_i\sim20}{cm$^{-3}$}, all in agreement with the previous orbit of MAVEN \cite{Burne2021}, except for slightly higher plasma temperatures. The magnetosonic Mach number $M_\text{ms} = V_\text{sw}/\sqrt{V_{a}^2+V_s^2}$ (with Alfvén and sound velocities $V_a$ and $V_s$) had a value around 6 (supercritical shock), whereas the solar wind ion plasma-$\beta$ reached to about $1.3$. This indicates that the solar wind conditions remained quite steady over the 3 hours between the two quasi-perpendicular crossings, with no CME or other solar transients effects disturbing the Martian plasma environment.
	
	\begin{figure}
		\noindent\includegraphics[width=\textwidth]{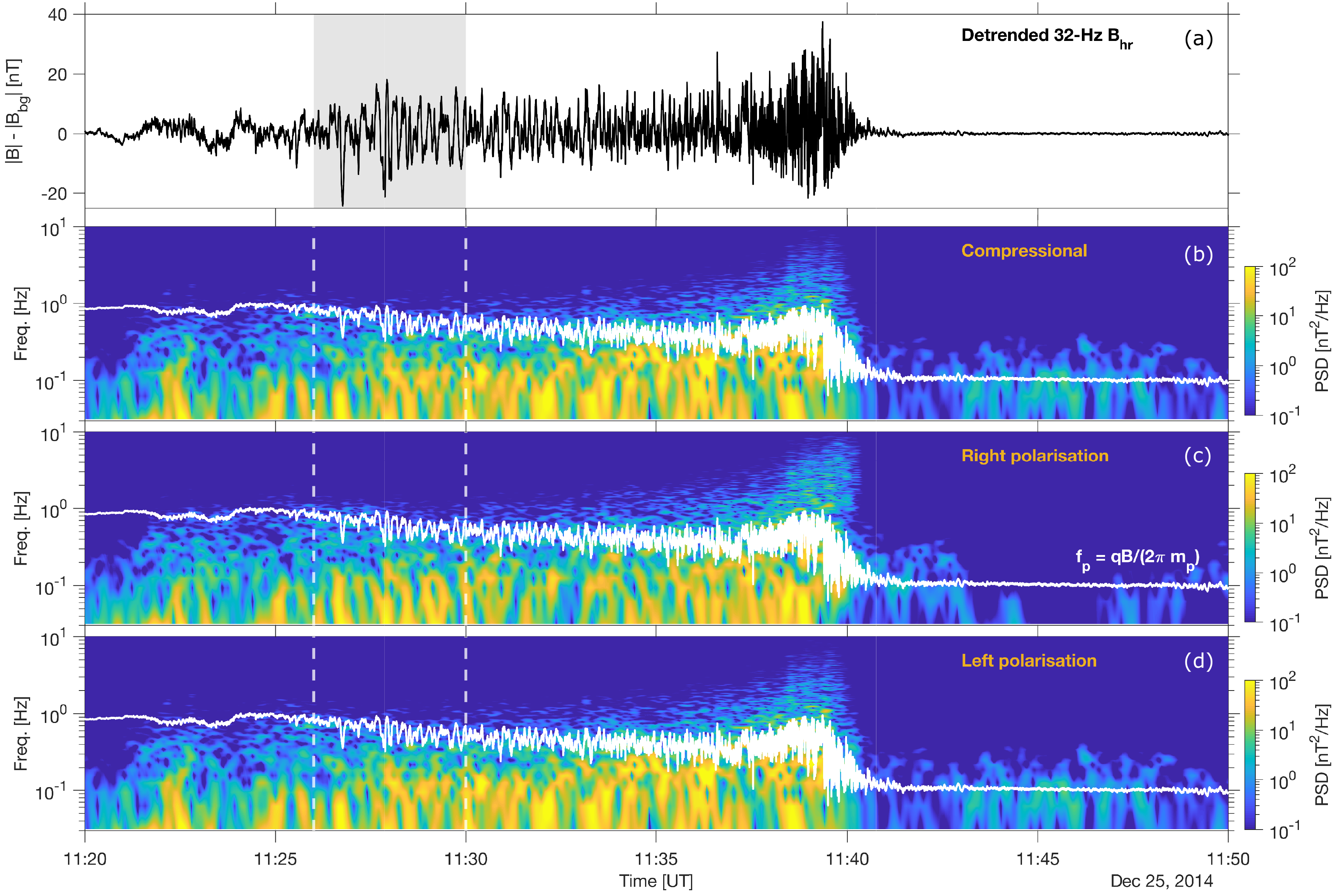}
		\caption{High-resolution magnetic power spectral density (PSD) as recorded by MAVEN/MAG, 25 December 2014 between 11:20 and 11:50\,UT (see Figure\ \ref{fig:MMevent1_contextGeneral} for the same time interval). (a) Module of the detrended $32\text{-Hz}$ magnetic $|\v B|-|\v B_\text{bg}|$, (b) compressional PSD, (c) right-handed polarisation PSD and (d) left-handed polarisation PSD. The plasma frequency calculated as $f_p = \frac{q |\v B|}{2\pi\,m_p}$ is superimposed as a white continuous line on panels b, c and d. The two vertical dashed lines and grey zone encompass the interval where clear mirror mode signatures were found, as explained in the text.
		}
		\label{fig:MMevent1_contextPSD}
	\end{figure}	
	
	The wave content for the full interval is shown in Figure\ \ref{fig:MMevent1_contextPSD}, where the power spectral densities (PSD) for the compressional, right-handed and left-handed polarisation terms are plotted using the \qty{32}{-Hz} magnetic field data. First, the matrix of passage $\v M_\text{MFA}$ from MSO coordinates to the MFA reference frame is calculated as in Equation\,(\ref{eq:app:matrixPassageMFA}) (\ref{appendix1}), using the background magnetic field $\v B_\text{bg}$ direction as the $z$ direction of the MFA system. Then the MSO $\v B$-field vector is rotated into MFA so that the two first directions are perpendicular to $\v B_\text{bg}$, noted $B_{\perp1}$ and $B_{\perp2}$, and the third one is parallel to $\v B_\text{bg}$, noted $B_{||}$. The compressional component, left-handed and right-handed polarisations are thus defined as:
\begin{linenomath}
\begin{align}
		B_C &= B_{||} &\text{Compressional}\\
		B_R &= \left(B_{\perp 1} + i B_{\perp 2}\right) / 2 &  \text{Right-handed polarisation}\\
		B_L &= \left(B_{\perp 1} - i B_{\perp 2}\right) / 2 &  \text{Left-handed polarisation}
	\end{align}
\end{linenomath}
where $i$ is the imaginary unit. Cross-spectrum analysis is performed with a FFT Welch method to evaluate the PSD of each component \cite{welch_use_1967}, using a moving window of $1024$ consecutive points (\qty{32}{s}) for a shift of $64$ points (\qty{2}{s}).
  
Most of the wave power is observed at frequencies much lower than the local cyclotron frequency, assuming only protons throughout, and, additionally, at frequencies lower than the solar wind proton cyclotron frequency (at about \qty{0.1}{Hz}). In the upstream region between 11:43 and 11:50\,UT, the signal appears mostly compressive with no right-handed polarisation component, which is consistent with the presence of proton cyclotron waves \cite{romanelli_proton_2016}. Downstream of the shock, the wave power is more evenly redistributed which suggests no clear relation between the upstream turbulence and the turbulence in the magnetosheath.

	\subsection{In-depth analysis}\label{sec:inDepthAnalysis}
	
\begin{figure}
		\noindent\includegraphics[width=\textwidth]{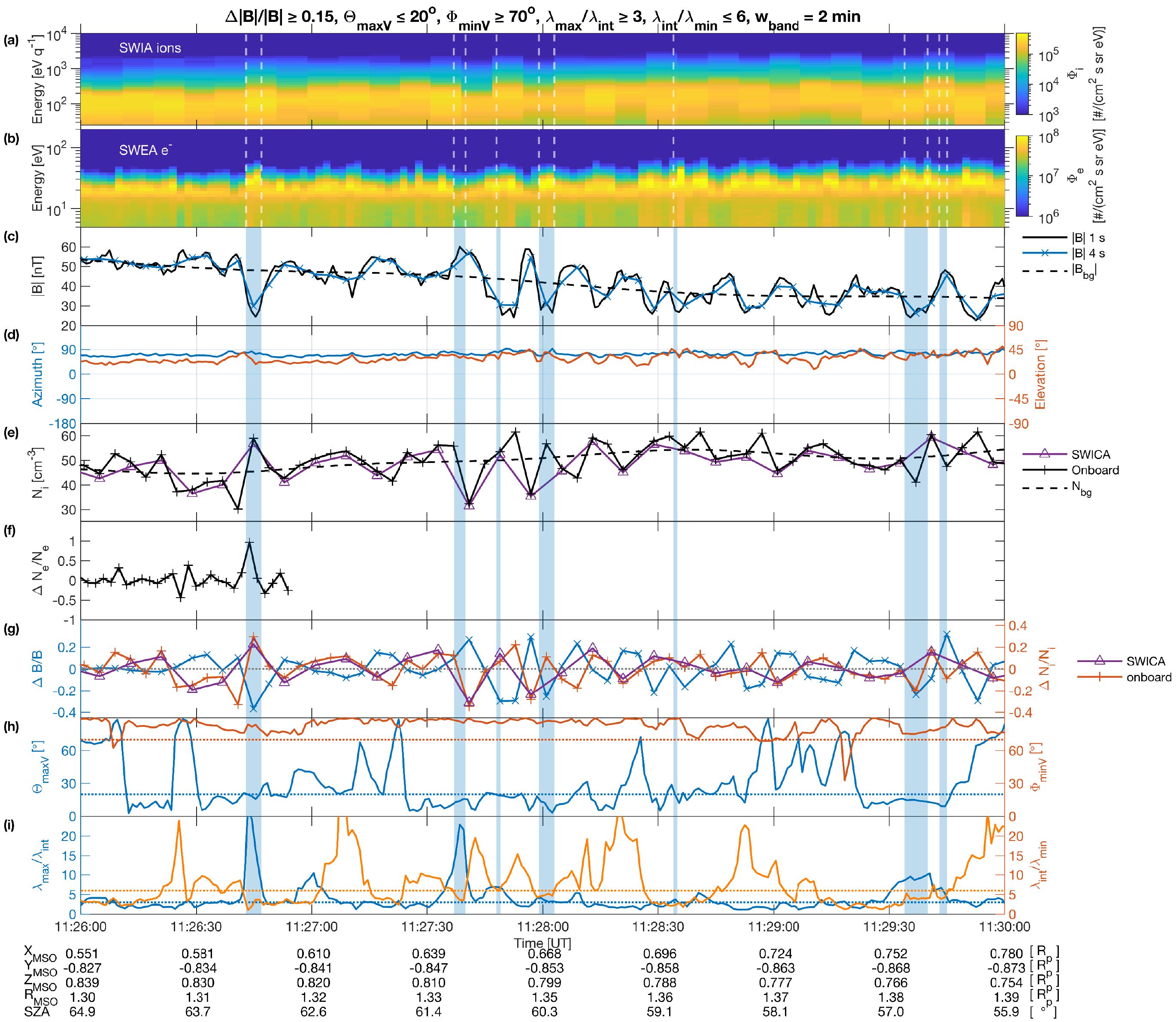}
		\caption{Mirror mode candidate structures seen with MAVEN on 25 December 2014. (a) Omnidirectional ion flux spectrum measured by SWIA and (b) omnidirectional electron flux spectrum measured by SWEA. During the interval, the field of view of SWIA did not encompass the magnetic field direction. (c) $B$-field data measured by MAG, (d) magnetic azimuth and elevation angles, (e) ion density measured with SWIA (onboard and SWICA modes), superposed with the average ion density $N_{\rm bg}$ (low-pass Butterworth filter applied to SWICA densities), (f) electron density fluctuations $\Delta N_e/N_e$ measured with SWEA (forward differences from integrated onboard energy spectra), (g) magnetic field and ion density fluctuations $\Delta B/B_\text{bg}$ (left axis, blue) and $\Delta N_i/N_i$ (right axis, with onboard data in orange and SWICA data in purple), (h) angles $\Theta_\text{maxV}$ and $\Phi_\text{minV}$ between the background magnetic field and the MVA directions (maximum and minimum variance directions), (i) eigenvalue ratios (maximum to intermediate, and intermediate to minimum). Horizontal dotted lines in panels h and i represent the thresholds corresponding to each parameter (colour-coded). Vertical light blue areas mark the identified intervals with candidate mirror mode structures fulfilling the $B$-field-only criteria of Table\ \ref{tab:Bcriteria}. Coordinates ($X_{\rm MSO}, Y_{\rm MSO}, Z_{\rm MSO}$, planetocentric distance $R_{\rm MSO}$), expressed in units of planetary radius $R_p$,  as well as the solar zenith angle (SZA) are indicated for the relevant time stamps.
		}
		\label{fig:MMdetection}
	\end{figure}

	In Figures\,\ref{fig:MMevent1_contextGeneral} and \ref{fig:MMevent1_contextPSD}, two vertical dashed lines highlight one interval between 11:26 and 11:30\,UT where fluctuations lodged in against the MPB appear to contain MM-like structures. This interval is shown as a zoom-in in Figure\ \ref{fig:MMdetection}, which displays ion/electron energy spectra (panels a and b), magnetic field $|B|$ (panel c), and ion densities $N_i$ and electron density variations $\Delta N_e/N_e$ (panels e and f). The resolution of the ion spectrum and calculated moments for the SWICA mode shown here is \qty{8}{s}, whereas the onboard moments have a resolution twice as good, i.e., \qty{4}{s}. SWICA and onboard moments agree rather well during this interval, with fluctuations closely following each other. The $2$-s resolution electron densities are unavailable after 11:26:45\,UT; electron density estimates calculated from the electron distribution function have afterwards a reduced cadence of \qty{8}{s}, which is too large to be of use for the typical width of the dips found in the magnetic field data. Hence, only the maximum resolution available is displayed; moreover, because these electron densities are not fully calibrated, only relative variations, calculated here as a forward difference, are shown. 
	
	Candidate MM detections using the $B$-field only criteria of Table\ \ref{tab:Bcriteria} are highlighted as blue-shaded rectangles. Skewness and excess kurtosis of the interval 11:26--11:30\,UT are $-0.20\pm0.16$ and $-1.09\pm0.32$, respectively, indicating an interval containing mostly dips and a heavy tail distribution with large fluctuations. The first large dip in $B$-field around 11:26:45\,UT coincides with an increase in both electron and ion fluxes, flux energies and densities, an observation that seems accurate for all other detected dips in the interval, except the last one around 11:29:40\,UT for which the SWICA poor spectral resolution is inconclusive, possibly indicating the slow transition from dips to peaks in the MM $B$-field structure (see \citeA{joy_mirror_2006} for a clearer example of this effect). This is confirmed by the fact that when shifting the interval window by increments of \qty{+30}{s} (11:26:30--11:31:30\,UT, etc.), the skewness, initially negative, steadily increases to become positive for a shift of \qty{+1}{min}. Throughout these periods, both magnetic field azimuth and elevation appear very stable (panel d), which is consistent with the presence of MM structures. Looking more closely into the magnetic field measurement (panel c), it appears oscillations in the total $B$-field take place starting around 11:27:30\,UT, with a quasi-period of \qty{\sim9.3}{s}, as determined by a Fourier analysis of the smoothed detrended signal. Several of these peaks or dips are not captured by the automatic detection algorithm.
	
	Figure\ \ref{fig:MMdetection}g,h,i displays the magnetic field MM analysis and plasma density variations during the interval, with the associated MM candidate structures (vertical blue rectangles) detected in the magnetosheath by the $B$-field-only criteria of Table\ \ref{tab:Bcriteria} (column `initial values'). Density variations calculated either from the onboard moments and from the SWICA mode exhibit very similar variations throughout the interval as shown in panels (e) and (g).
	The detected candidate MM all present a clear $B-N_i$ anticorrelation as demonstrated with the relative $B$-field and density variations $\Delta B/B$ and $\Delta N_i/N_i$ (panel g). The gain in temporal resolution (\qty{4}{s}) from the onboard ion datasets allows the ion density variations to follow even more closely those of the magnetic field throughout the interval. A supplementary indicator is given for the first MM detection (dip) around 11:26:45\,UT by SWEA's $\Delta N_e/N_e$ at a resolution of \qty{2}{s}, which allows this \qty{8-9}{s}-long structure to be probed with almost $4$ electron density data points. The $B-N_i$ and $B-N_e$ anticorrelations for this event are striking, and represent the clearest and most unambiguous MM signature detected to date at Mars. The consistency found between onboard and manually calculated moments, as well as the electron density data for this event, gives confidence for the rest of the interval. Moreover, it suggests that possibly every period with increased electron and ion fluxes, energies and densities with respect to the background levels in Figure\ \ref{fig:MMdetection} are MMs proper. The only exceptions are maybe towards the end of the interval, where the two last structures display a very rough anticorrelation between ion density and magnetic fields, with eigenvalue criteria only just fulfilled, in contrast to the previous structures. When studying in more detail the electron spectrograms at those times (panel b), electron fluxes and magnetic field intensities do appear anticorrelated, giving more confidence that these are indeed MMs too. 
	
	Out of a measurement window of \qty{240}{s}, \qty{21}{s} fulfilled the MM criteria above, resulting in $4$--$5$ separate structures if two structures are separated by at least \qty{8}{s} (that is, half of the MVA window and, also, the temporal resolution of the SWICA moments in this region), as in \citeA{volwerk_mirror_2016}. The $B$-field-only algorithm captured many prominent MM structures that can be identified by eye, but discarded those that are deemed ambiguous. Moreover, because the magnetic field background level seems to cut through the middle of the oscillations in the second part of the interval, several promising structures were missed altogether.
	
	Uncertainties on the MVA directions and magnetic field components were evaluated following the prescriptions of \citeA{sonnerup_minimum_1998}. During the period concerned, the mean statistical error on the $B$-field magnitude along the maximum variance direction was less than \qty{5}{nT}, that is, about $10\%$ and below of the background field. Hodograms of the $B$-field variation in the MVA orthogonal planes (not shown) displayed clear linearly polarised signatures. This is giving additional confidence on the validity of the detections performed. For a discussion of MVA and how it is applied to the detection of ULF waves, the reader is for example referred to the review of \citeA{volwerk_multi-satellite_2006}. 

\begin{figure}
		\noindent\includegraphics[width=0.7\textwidth]{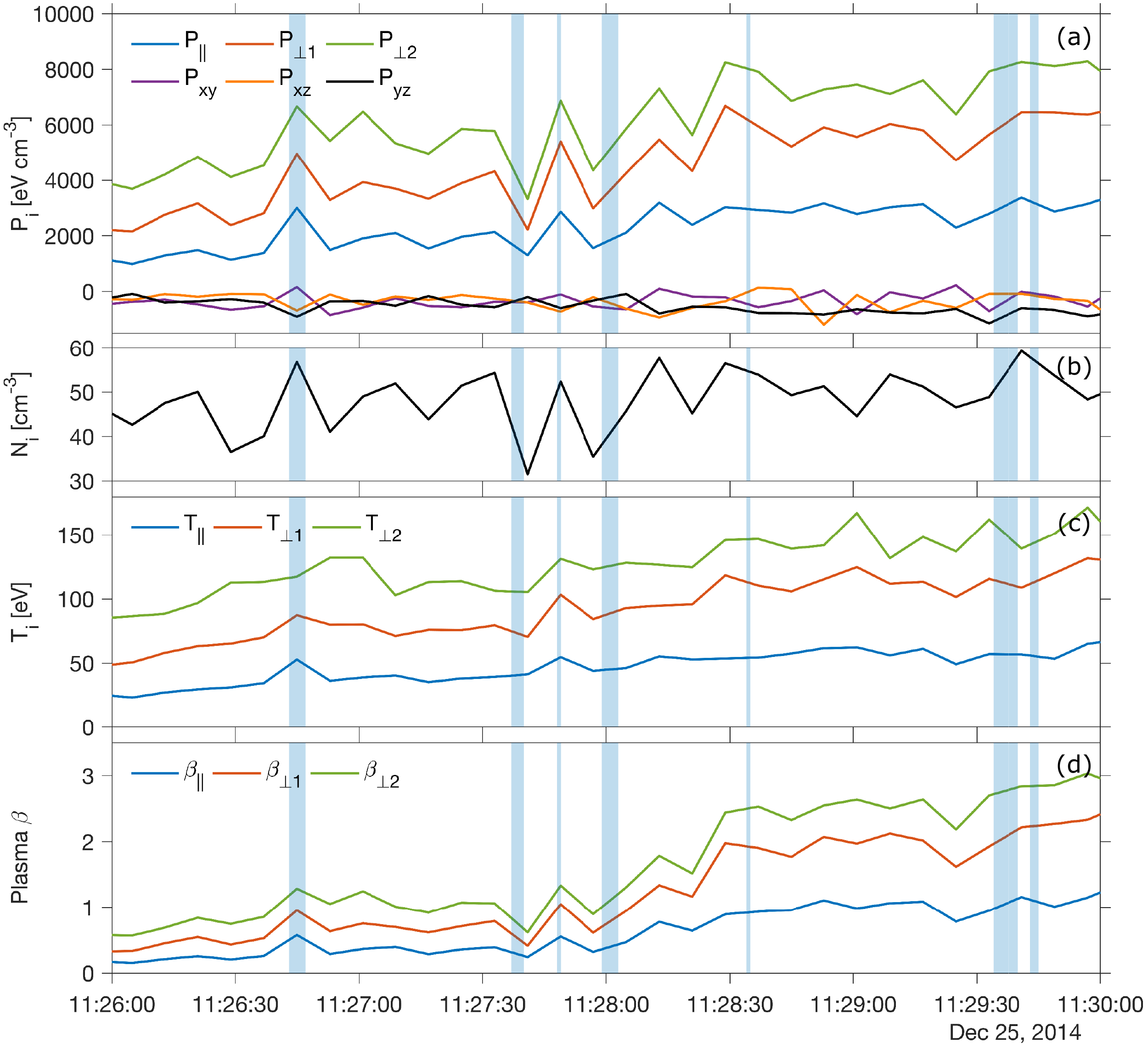}
		\caption{Retrieved moments of the ion distribution function, SWIA Coarse Archive mode and derived quantities necessary to calculate the MM instability criterion (MMIC). (a) parallel and perpendicular pressures, (b) ion density, (c) parallel and perpendicular temperatures, (d) parallel and perpendicular plasma-$\beta$. MM candidates detected with $B$-field-only criteria are highlighted as vertical light blue areas. 
		}
		\label{fig:moments}
	\end{figure}

    Let us now investigate further the points raised in Section\ \ref{sec:criteriaValidation}, i.e., magnetic field direction and instability criterion MMIC. First, during the interval, the magnetic field azimuth and elevation angles do not change significantly, with less than $15\deg$ rotation (standard deviations $\leq 7.5\deg$). More precisely, angular changes across each MM-like structures are less than $5\deg$ on average, in excellent agreement with the theoretical and observational expectations of MMs \cite{tsurutani_magnetosheath_2011}.  
    Second, the MMIC makes use of the parallel and perpendicular temperatures which are calculated from the pressure tensor as explained in Section\ \ref{sec:criteriaValidation} and \ref{appendix2}, with the MFA system based on $\v B_\text{bg}$ as defined in \ref{appendix1}.
    Results for $P_{||}$, the two perpendicular directions $P_{\perp1}$ and $P_{\perp2}$, as well as corresponding temperatures and plasma-$\beta_{||}$ and $\beta_\perp$ are shown in Figure\ \ref{fig:moments}. The diagonalisation method yields off-diagonal components of the pressure tensor that are small in comparison to the diagonal elements, with larger fluctuations seen when entering the candidate MM structures, especially the one around 11:26:45\,UT. Clear anisotropies in pressure (and hence temperatures, although their variations are in comparison somewhat dampened by the abrupt variations seen in the densities) are seen, with the two perpendicular components larger than the parallel component by a factor $2$ or so. Because the mean magnetic field vector was in the blind sector of SWIA's limited FOV, the parallel components in pressure, temperature and plasma-$\beta$ are all expected to be lower estimates. This in turn implies that our calculated anisotropies $T_\perp/T_{||}$ averaging $2.2$ over the studied interval. are upper estimates (see Figure\ \ref{fig:app:Pressure}c). However, our retrieved anisotropies are certainly not anomalous as values of $2$ or more are common in the Martian magnetosheath \cite{halekas_flows_2017}.

\begin{figure}
		\noindent\includegraphics[width=\textwidth]{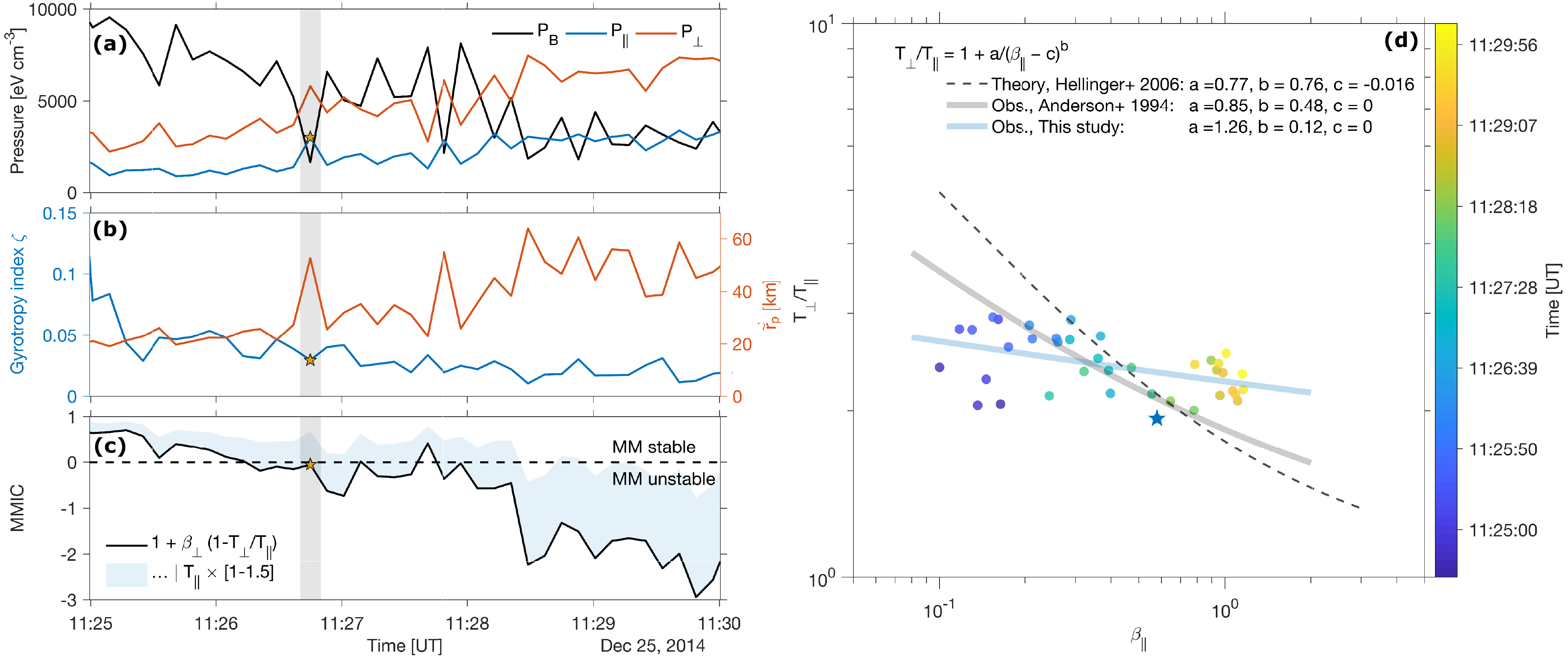}
		\caption{Parallel and perpendicular ion pressures $P_{||}$ and $P_\perp$ (\qty{8}{s} resolution) and magnetic pressure $P_B$ at \qty{1}{s} (panel a), gyrotropy index $\zeta$ (panel b, left axis, see Equation\,\ref{eq:app:gyrotropy}) and equivalent spatial scale commensurable to a fluid-like proton gyroradius $\Tilde{r_p}=m_p\,V_\perp/q|B|$ (panel b, right axis), mirror mode instability criterion (panel c) and $T_\perp/T_{||}$ as a function of plasma $\beta_{||}$ (panel d). Because of the difficulty in estimating the two perpendicular directions, the maximum eigenvalue is retained in these calculations so that $T_{\perp} = \max_j(T_{\perp j})$. The grey-shaded zone on the left highlights the times when the clearest MM structure is detected throughout the interval: a star marks the position of the middle of the zone at the resolution of SWIA. An inverse regression in $T_\perp/T_{||} = 1 + a/\left(\beta_{||}-c\right)^b$ is given as a comparison to previous works, theoretical \cite{hellinger_solar_2006} or observational \cite{anderson_magnetic_1994}. 
		}
		\label{fig:MMIC}
	\end{figure}

Figure\ \ref{fig:MMIC} shows the interplay between magnetic and ion pressures (parallel and perpendicular) (panel a). Magnetic pressure $P_B = |B^2|/2\mu_0$ and ion pressures, especially the perpendicular pressure $P_\perp$, appear to be rather well anticorrelated throughout the interval, at the resolution of SWIA. This is in agreement with results at comets \cite{mazelle_etude_1990,mazelle_ultra_1991}. 

In order to give a measure of the non-gyrotropy of the pressure tensor, the quantity $\zeta$ (Equation\,\ref{eq:app:gyrotropy}, \ref{appendix2}), which we refer to as ``gyrotropy index'' based on the off-diagonal elements of the pressure tensor following \citeA{swisdak_quantifying_2016}, is shown in panel (b), left axis. Any value of $\zeta\neq0$ is a departure from the normal distribution (for which $\zeta=0$), which is observed here throughout the interval, with $\zeta = 0.05-0.12$ (mild departure from perfect gyrotropy). Early in the interval the gyrotropy index is much larger: this can be ascribed to the presence of heavy pickup ions, confirmed by a cursory look at the mass spectra from STATIC (see also Figure\ \ref{fig:MMevent1_contextGeneral}), essentially unmagnetised early in their trajectory.
It is interesting to note that when entering the most prominent MM dip structure around 11:26:45\,UT, $\zeta$ decreases together with the magnetic field intensity; consequently, gyrotropy index and perpendicular pressures appear also anticorrelated, as can be seen later in the interval. This implies that in an MM structure, the off-diagonal elements of the pressure tensor tend to diminish with respect to the diagonal (parallel and perpendicular) elements. 

How does that compare with spatial scales probed by MAVEN throughout the interval? One way of evaluating this aspect is to define a quantity commensurable with a ``fluid-like'' gyroradius for core-distribution protons, i.e., an equivalent spatial scale calculated as $\Tilde{r_p} = m_p\,V_\perp/q|B|$, assuming most ions (the core ions) have a perpendicular velocity equal to $V_\perp$, the perpendicular bulk plasma velocity. In this way, $\Tilde{r_p}$ would be a rough estimate of the mean ion gyroradius for ions populating the maximum of the non-Maxwellian ion distribution function. The decrease in gyrotropy index $\zeta$ coincides thus with $\Tilde{r_p}$ (panel b, right axis) increasing by almost a factor $2$ in mirror mode 'dip' structures, because of a corresponding decrease of the magnetic field intensity. One tentative interpretation in a kinetic sense could be that ions with larger gyroradii (and potentially seeping into the mirror loss cone of the `magnetic bottle') escape, eventually tending towards a more gyrotropic distribution. This effect could become especially important for smaller MMs or MHs. We note also that $\Tilde{r_p}$ and the local thermal proton gyroradius, determined from $T_\perp = P_\perp/N_i$, have very similar values during the interval considered.

The MM instability criterion MMIC is shown in Figure\ \ref{fig:MMIC}c. The region crossed by the spacecraft and containing the train of MMs is interpreted as being marginally unstable to the generation of MM structures, with ${\rm MMIC}$ moving from positive values before 11:26\,UT ($0\leq{\rm MMIC}<1$) to negative values afterwards. Significantly negative values (${\rm MMIC}\leq-1$) start to occur starting 11:28:30\,UT and onwards, coinciding with the start of the quasi-periodic $B$-field signal. However, because of the likely overestimate of $T_\perp/T_{||}$, the exact time where the MMIC may cross the threshold line ${\rm MMIC}=0$ is likely inaccurate. In order to investigate how changes in $T_{||}$ may modify this interpretation, we decided to increase by increments of $10\%$ the value of $T_{||}$ up to a $50\%$ overall increase, which we expect to be much larger than the uncertainty on the temperature retrievals (see discussion in Section\ \ref{sec:criteriaValidation}), hence creating an ``envelope of confidence'' for our interpretation. This is illustrated as a light blue region above the nominal level in black in Figure\ \ref{fig:MMIC}c. This increase shifts the curve upwards towards the stability line at ${\rm MMIC}\sim0^-$, making the plasma reach marginal stability almost all the time during the interval. 
At 11:26:45\,UT, in the middle of the most prominent MM structure found (vertical light grey area on the figure), the derived MMIC  remains very stable and close to the threshold. Combined with predominantly MM-unstable conditions at play for times after around 11:29\,UT, this would imply that the region crossed by MAVEN at that time is not the source region of these MMs and should be found upstream of the spacecraft. Owing to their duration (around \qty{10}{s}, i.e., of the order of \qty{1000}{km} in size for a lower-estimate plasma velocity of \qty{100}{km/s}) and possibly advanced development stage (mixed presence of dips and peaks), this source region may be up to several hundreds of kilometres upstream in size, and not necessarily in the path of the spacecraft.

In Figure\ \ref{fig:MMIC}d, the temperature anisotropy $T_\perp/T_{||}$ is shown as a function of proton plasma-$\beta_{||}$ and time (colour code), with a Levenberg-Marquardt fit of the form: 
\begin{linenomath}
\begin{align}
    T_\perp/T_{||} = 1 + \frac{a}{\left(\beta_{||}-c\right)^b}, \label{eq:anisotropyFit}
\end{align}
\end{linenomath}
as in \citeA{fuselier_inverse_1994} and \citeA{hellinger_solar_2006} 
to investigate the marginal stability condition with respect to MM and ion cyclotron instabilities. The modified MMIC criterion translates to $[a=0.77,\,b=0.76\,c=0.016]$ \cite{hellinger_solar_2006}, with any point on the right of that line denoting MM-unstable conditions. In the interval of study, $a = 1.26$ and $b = 0.12$ with $c=0$, which is in contrast with previous studies at Earth of the proton anisotropy: \citeA{anderson_magnetic_1994} obtained values $a = 0.85$ and $b = 0.48$ whereas \citeA{genot_mirror_2009-1} found $a = 0.47$ and $b = 0.56$ ($c=0$ in both cases). A flatter inverse correlation with $\beta_\perp$ ($a = 1.41$, $b = 0.09$, not shown) was found in the study interval. This in turn is closer to the values found by \citeA{fuselier_inverse_1994} ($a=1.40, b=0.26$) for He$^{2+}$ ions, also in the Earth's magnetosheath. Such differences are not surprising since (i) our present analysis does not discriminate between protons and other ions (including alpha particles) in the ion distributions, and (ii) parallel temperatures are likely underestimated because of the FOV of SWIA. Moreover, we limited ourselves here to one event spanning a few minutes only. A statistical study of the temperature anisotropy behaviour with respect to the plasma-$\beta$, using the full MAVEN dataset, is left for another study.

\subsection{Associated scales and validation of $B$-field criteria}\label{sec:reanalysis}

\begin{figure}
		\noindent\includegraphics[width=\textwidth]{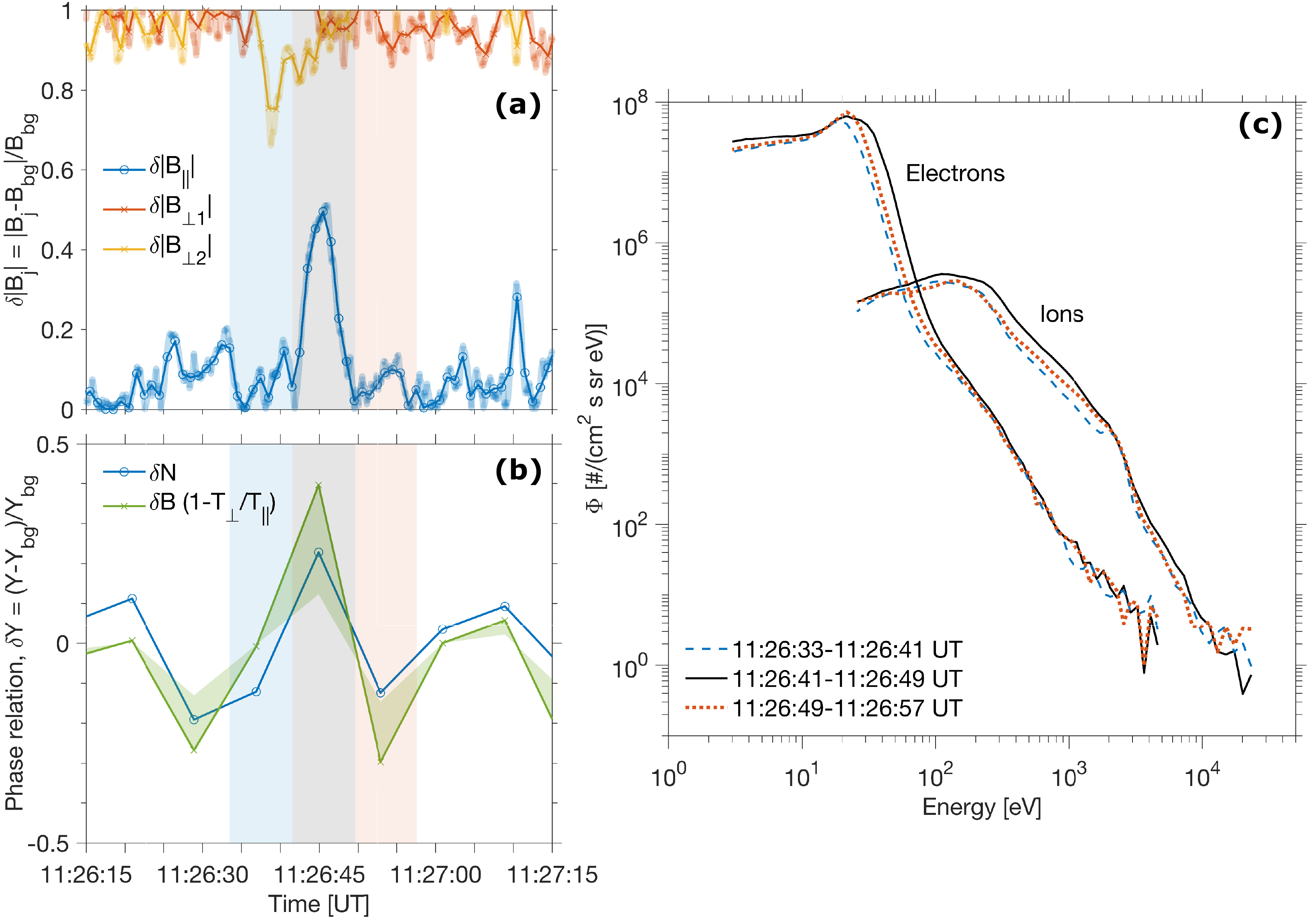}
		\caption{Clearest MM structure detected around 11:26:45\,UT and lasting almost \qty{9}{s}. 
		(a) Normalised parallel and perpendicular components of the magnetic field fluctuations $\delta |B_j|$ $=$ $|B_j$ $-$ $B_\text{bg}|/B_\text{bg}$ at \qty{1}{Hz} (solid lines) and \qty{32}{Hz} (thicker semi-transparent lines), with $j$ $=$ $(||,\text{$\perp1$},\text{$\perp2$})$ directions. 
		(b) Phase relation where the two sides of Equation\,(\ref{eq:phaseDiff}) are compared. Normalised fluctuations $\delta N$ $=$ $(N-N_\text{bg})/N_\text{bg}$ for ions (SWIA density measurements) and $\delta B \left(1-T_\perp/T_{||}\right)$ (with $\delta B$ $=$ $(|\v B|$ $-$ $B_\text{bg})/B_\text{bg}$) were calculated using a low-pass Butterworth filter to estimate the background contributions $N_\text{bg}$ and $B_\text{bg}$. The light green-shaded region illustrates an increase in $T_{||}$ of up to $50\%$ to evaluate how sensitive the results are to an underestimation of $T_{||}$ (see text for explanation). 
		(c) Particle differential electron and ion fluxes inside and outside the MM structure.
		Black lines correspond to the mean fluxes inside of the structure (grey zone on panels a and b) whereas the blue and red lines correspond to mean fluxes \qty{8}{s} (= resolution of SWICA) outside of the structure (light blue and red zones on panels a and b). Notice that the zones do not start exactly at a full second since the original temporal width of the MM was determined by an interpolation of the \qty{1}{-Hz} total $B$-field intersecting the background field level $B_\text{bg}$.
		}
		\label{fig:fluxesPhase}
	\end{figure}

A reanalysis of the current interval is now possible on the basis of the plasma measurements and the high confidence level in the density variations. The temporal width of each structure is evaluated by remarking that the first crossing between the background magnetic field level on either side of the detection constitutes the start and end points of the structure. It is clear that each event has a different scale in the parallel and perpendicular directions, as remarked by \citeA{zhang_characteristic_2008}: this is seen in Figure\ \ref{fig:fluxesPhase}a for the clearest event around 11:26:45\,UT, when comparing the parallel and cross-field magnetic fluctuation components normalised to the background field, $\delta \left|B_{||}\right| = \left|B_{||}-B_\text{bg}\right|/B_\text{bg}$ and $\delta \left| B_\perp\right|$.
 
For this particular event, one can specifically check for the species-by-species phase relationship between density and the magnetic perturbations, which we apply here to the ion measurements. It is approximately, in the hypothesis of a cold component of the plasma, in this case the cold photoelectrons \cite{hasegawa_drift_1969,mazelle_ultra_1991}:
\begin{linenomath}
    \begin{align}
    	\frac{\Delta N_i}{N_i} = \left( 1-\frac{T_\perp}{T_{||}} \right) \frac{\Delta B}{B}. \label{eq:phaseDiff}
    \end{align}
\end{linenomath}
Equation\,(\ref{eq:phaseDiff}) formalises that for a temperature anisotropy in the perpendicular direction ($T_\perp/T_{||} > 1$), the density is expected to be out of phase with the magnetic field. In Figure\ \ref{fig:fluxesPhase}b, we show the comparison between the left-hand side and the right-hand side of the equation, with $\delta N_i = \Delta N_i/N_i$. In order to evaluate the uncertainty in the parallel temperature (derived from SWIA's limited field of view), the shaded green region denotes a $50\%$ variation of $T_{||}$ ($1-1.5\times T_{||}$). Although the $\delta B$ term has larger variations in the time span considered, the agreement, both in shapes and amplitudes of variations, is striking: it quantitatively confirms the anticorrelation between density and magnetic field fluctuations and the ion density. Similarly, \citeA{mazelle_ultra_1991} found that Equation\,(\ref{eq:phaseDiff}) was also accurate for the low-frequency MM-like signatures found at comet 1P/Halley.

To try to further characterise the particle content of this MM, the mean electron and ion omnidirectional fluxes measured by SWEA and SWIA are shown in Figure\ \ref{fig:fluxesPhase}c where mean fluxes were derived successively inside and outside the structure (see details in the caption). As already remarked from Figure\ \ref{fig:MMdetection}, the inside of the MM structure corresponds to a distinct increase in both peak energy and flux with respect to the outside, both for electrons and for ions. Interestingly, the fluxes on either side of the MM are similar, suggesting that the surrounding plasma may well have very similar characteristics. This paints the picture of a magnetic bottle containing a relatively energetic and dense ion population, and drifting with the surrounding cooler and diluted plasma.
 
Reanalysis of all the MM structures present in the 11:26--11:30\,UT interval proceeded as follows. First, we manually select the location and temporal width of $9$ ``dip'' MMs for simplicity, assuming that their peak counterparts are part of the same structure, by comparing the $|B|$-field variations with the onboard moment and SWICA density variations. Their average width is \qty{8.7\pm2.1}{s}, with an average dip-to-dip period of \qty{\sim9.3}{s}. Four structures last up to \qty{10}{s} whereas the shortest around 11:28:28\,UT last only about \qty{5}{s}. Assuming a plasma bulk speed of about \qty{150}{km/s} (mostly in the perpendicular direction to the magnetic field as shown by SWIA), and that the spacecraft is at rest with respect to the plasma, the size of the MM structures varies between $750$ and \qty{1500}{km} (a large fraction of the planetary radius, $0.22-0.44\,R_p$), 
 which is much larger than the equivalent spatial scale of the peak-distribution ions in this region (\qty{\Tilde{r_p}\sim35}{km}, i.e., MM sizes of the order of $\sim20$--$40\,\Tilde{r_p}$, see Figure\ \ref{fig:MMIC}b). By comparison, we can estimate the thermal proton gyroradius, noted $r_p=\sqrt{2k_b T_\perp/m_p}/q|B|$, in the possible source regions, either in the solar wind or in the immediate wake of the quasi-perpendicular shock. In the solar wind upstream of the shock, around 11:45\,UT, the perpendicular proton temperature is \qty{6.5\times10^4}{K} (i.e, $T_\perp=P_\perp/N_i\sim5.6$\,eV as measured on average from the SWIFA mode -- a textbook value, see \citeA{slavin_solar_1981}) and the upstream $B$-field magnitude was \qty{\sim6.6}{nT}. This corresponds to a solar wind proton thermal gyroradius $r_p^\text{sw}$ of the order of \qty{50}{km} (\qty{\sim0.015}{$R_p$}), implying that MM sizes vary from $\sim15$ to $30\,r_p^\text{sw}$ throughout this interval. Such an estimate, albeit significantly smaller, is in line with similar considerations at Venus made by \citeA{zhang_characteristic_2008}, who deduced MHs to have sizes of $40-100\,r_p$ in the shape of a prolate spheroid. Similarly, just behind the shock around 11:37\,UT where the magnetic field intensity is lower than at the location of the MMs, $T_\perp\sim 150$\,eV, and $|B|\sim 24$\,nT, resulting in a magnetosheath thermal gyroradius $r_p^\text{ms}\sim 75$\,km, i.e., the observed MMs deeper in the sheath have scales $\sim10\text{--}20\,r_p^\text{ms}$. The typical sizes given here are in agreement with previous studies (see Table\ 2 of \citeA{tsurutani_magnetosheath_2011}).

\begin{figure}
		\noindent\includegraphics[width=\textwidth]{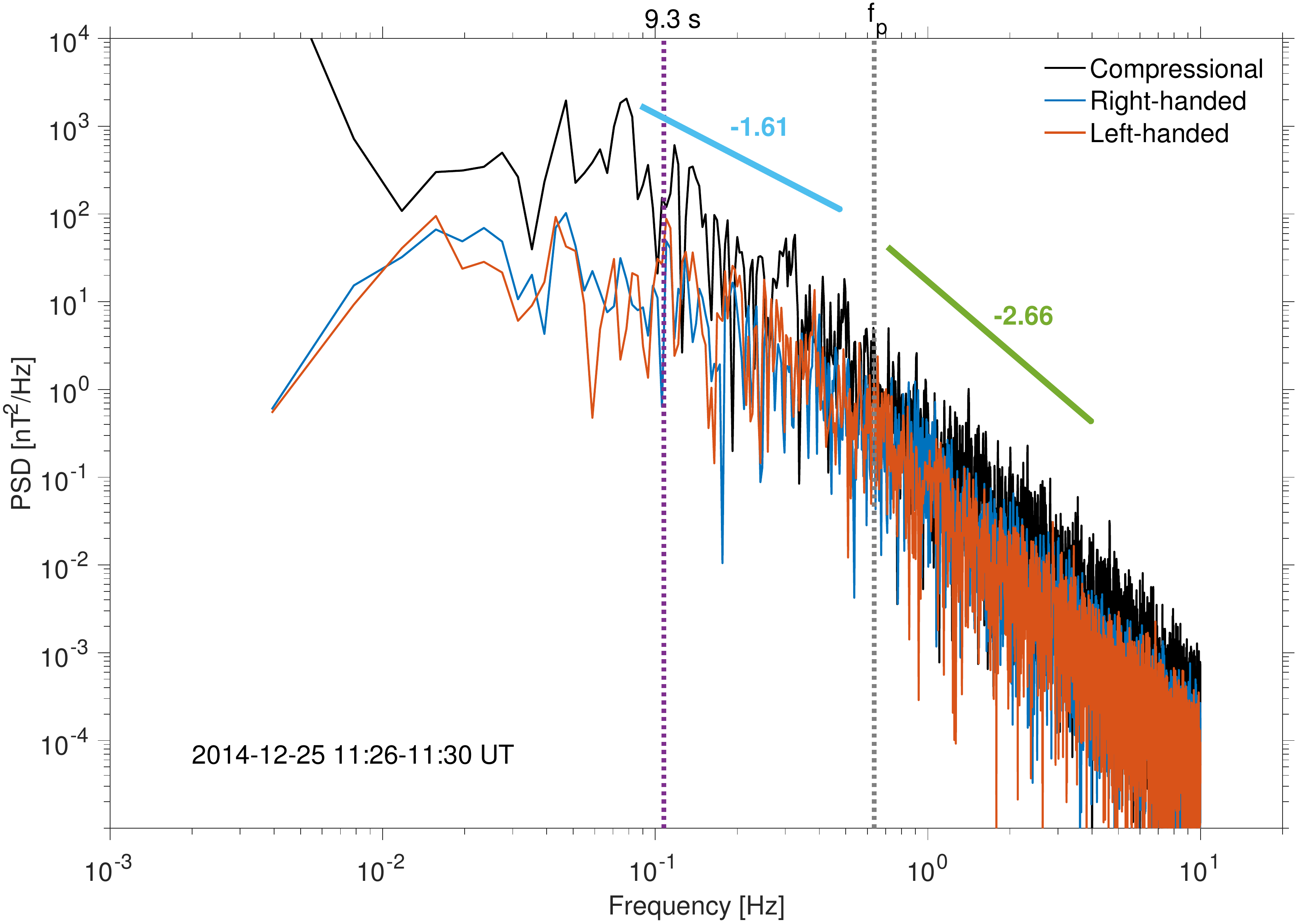}
		\caption{Welch-estimated power spectral densities (PSD) for the magnetic compressional, right-handed and left-handed polarisations, using the \qty{32}{Hz} magnetic field data, on 25 Dec. 2014 11:26--11:30\,UT. The vertical line indicates the approximate period of \qty{9.3}{s} (\qty{f_\text{mm}$ $\sim$ $ 0.017}{Hz}) found for MM 'dip' structures with the revised $B$-field only criteria. Tendencies towards Kolmogorov-like turbulence (spectral index $-1.61$ $\approx$ $-5/3$) and steepened spectra at the sub-ion scales ($-2.66$ $=$ $-8/3$) are shown for comparison. The mean proton gyrofrequency for the interval, \qty{\overline{f}_p = q|\v B|/2\pi\,m_p = 0.64}{Hz} is indicated.
		}
		\label{fig:PSD}
	\end{figure}

We also look into the magnetic field power spectral density (PSD) estimates of the compressional, right-handed and left-handed components during the full interval, computed with Welch's algorithm and a $50\%$ Hamming window overlap \cite{welch_use_1967}. This is shown in Figure\ \ref{fig:PSD}, with the mean MM period of \qty{9.3}{s} (\qty{0.017}{Hz}) appearing to almost coincide with a series of larger peaks in the compressional PSD, with the two circular polarisations being muted in comparison. During the interval considered, the mean proton gyrofrequency $\overline{f}_p$ was equal to \qty{0.64}{Hz}: above this limit, sub-ion scales start. Two main trends in the overall PSDs can be seen, with a Kolmogorov-type turbulence (spectral index $-1.61\sim -5/3$ in the fluid limit) steepening above the proton gyrofrequency (spectral index $-2.66=-8/3$). The mean period of the MMs lies thus in the Kolmogorov-like region of the spectrum, as expected from studies of the low-frequency wave turbulence in the Earth's magnetosheath \cite{sahraoui_ulf_2003,sahraoui_correction_2004} or from simulations \cite{hellinger_mirror_2017}. This is also in agreement with the study of \citeA{ruhunusiri_characterization_2017} who calculated turbulence spectra in the environment of Mars with MAVEN magnetometer data. They found on average spectral indices of the order of $-1.43$ and $-2.50$ close to the MPB, values which are close to our estimates. This has one main consequence for the evolution of our detected MMs: the environment in which these MMs are embedded has had time to evolve into a fully developed energy cascade.

\begin{figure}
		\noindent\includegraphics[width=\textwidth]{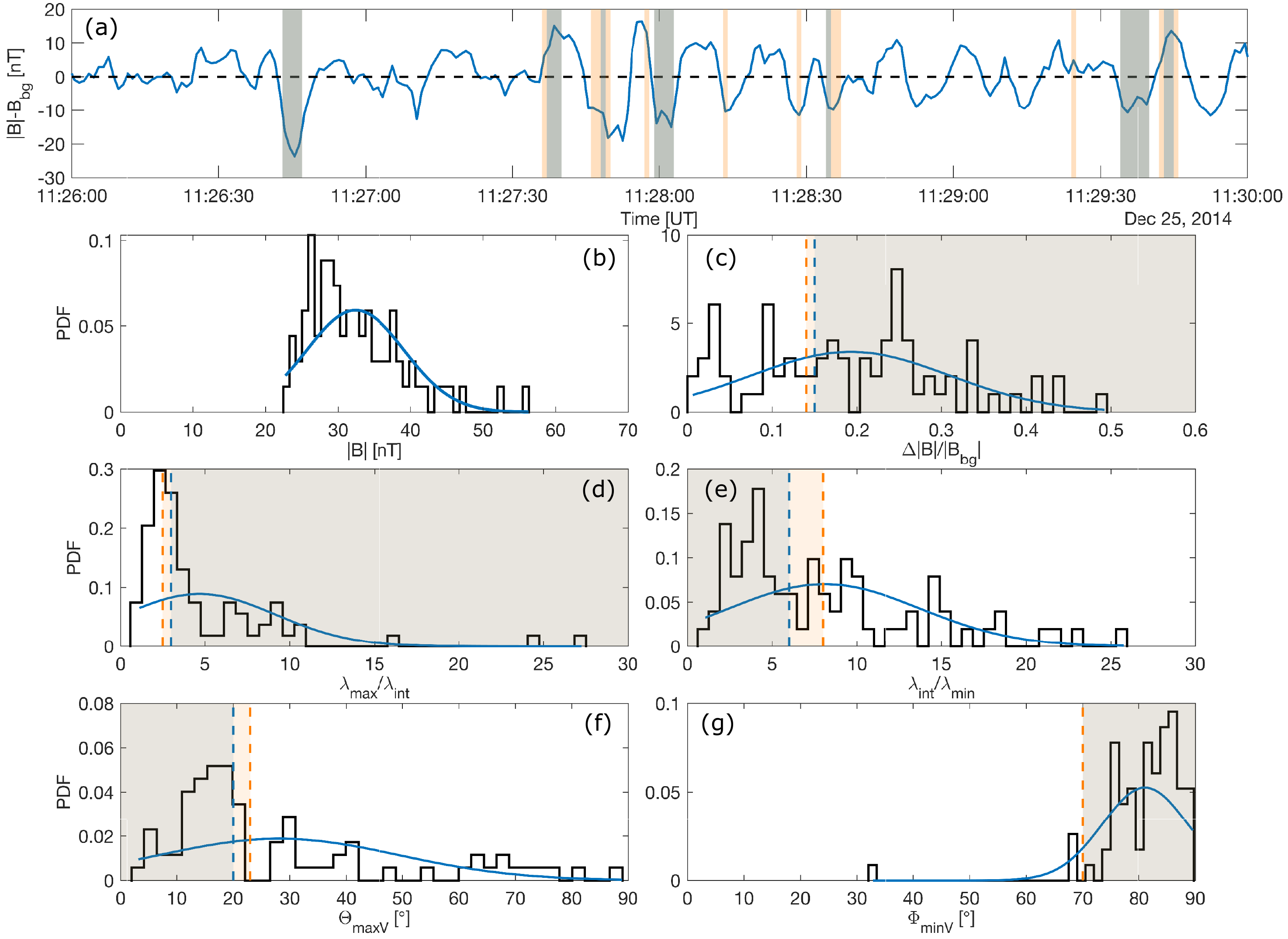}
		\caption{Probability density function (PDF) of $9$ manually picked mirror modes during the interval 11:26--11:30\,UT. These structures lasted in total \qty{78}{s} in this \qty{240}{s} interval. (a) Detrended total magnetic and detection of MMs. (b) Magnetic field intensities, (c) fluctuations $\Delta|B|/|B_\text{bg}|$, (d) eigenvalue ratio $\lambda_\text{max}/\lambda_\text{int}$ and (e) $\lambda_\text{int}/\lambda_\text{min}$, (f) and (g) angles between background field and maximum (respectively, minimum) variance $\Theta_\text{maxV}$ ($\Phi_\text{minV}$) directions. Initially retained ranges from Table\ \ref{tab:Bcriteria} are highlighted in transparent blue, whereas the revised criteria, determined from the PDFs within $1\sigma$ of the mean, yield the ranges in orange. Normal distribution fits are shown as blue lines. Out of $240$ points in the interval, the total number of revised measurement points was $33$ (initial criteria: $21$ points), thus equal to a total residence time in a MM structure of \qty{33}{s}. 
		}
		\label{fig:statistics}
	\end{figure}

Finally, manually picking the location and duration of MMs in our interval results in the statistics shown in Figure\ \ref{fig:statistics}, with $9$ structures totalling \qty{78}{s}, from which a revised set of $B$-field only criteria was extracted: this new set of criteria has the advantage of being unbiased with respect to other choices of criteria adopted at Venus and Earth. Criteria were on average relaxed ($\Delta|\v B|/|B_\text{bg}|$, $\lambda_\text{max}$ and $\lambda_\text{min}$, $\Theta_\text{maxV}$) 
to better take into account the characteristics of the detected MMs. 
To detect as many MMs as possible and limit false positive detections, we decided to keep within $1\sigma$ of the mean of the PDFs shown in Figure\ \ref{fig:statistics}, depending on the spread in values within the range. The revised detections are highlighted in orange, whereas the detections using the initial criteria are in light blue; common periods are in grey. The revised set of criteria is given in Table\ \ref{tab:Bcriteria}, yielding $7$ 'dip' structures and $3$ 'peak' structures, with $1$ potential outlier, totalling \qty{33}{s} out of a \qty{240}{-s} interval. Some of these peak structures appear to be nothing more than counterparts of dip structures, depending on the choice of the background field. Their location with respect to detrended $B$-field values, where MM structures have typical peak-to-peak amplitudes of \qty{20-35}{nT}, are also emphasised as areas in light orange on the figure.

Robust detections across an entire orbit may be difficult as small adjustments in angles and eigenvalue ratios can change the results, increasing or decreasing the number of potential candidates. One difficulty lies in the aforementioned mixed presence of peaks and dips in the study interval considered. Automatic detection applied to larger datasets may in this way benefit from considering smaller intervals of a few minutes on which to calculate the background field levels, depending on the skewness of the $B$-field distribution in that interval \cite{genot_mirror_2009-1,ala-lahti_statistical_2018}. However, we stress that any attempt at detecting MMs based on $B$-field only measurements will always result in false positive detections and that plasma data is necessary for a positive identification.

\section{Discussion}\label{sec:discussion}

Although magnetic field signatures in the magnetosheath of Mars were previously found consistent with the probable presence of MM waves \cite{bertucci_mgs_2004,espley_observations_2004}, no dedicated study has been possible until the advent of MAVEN and its full magnetic field and high-time resolution plasma suite of instruments. Consequently, we presented in this study the first in-depth characterisation of MMs in the Martian environment.

Where do these structures originate? The question of the origin and development of MM structures around an unmagnetised object such as Mars, Venus or around comets is of particular interest to the space physics community. Through competition with the left-hand Alfvén ion cyclotron mode (which heats the plasma), MMs consume the locally available free energy of the temperature or pressure anisotropy and may help in the thermalisation of the magnetosheath \cite{shoji_mirror_2009,soucek_properties_2008}. The source of the anisotropy can be twofold: (i) it may come from the intrinsic properties of the plasma in the wake of the quasi-perpendicular shock which heats and deflects the solar wind \cite{bale_quasi-perpendicular_2005,peng_kinetic_2015} or (ii) from the pickup ion process several proton gyroradii in the upstream solar wind \cite{Price1989,bader_proton_2019}. The first mechanism occurs at magnetised (Earth, Jupiter) and unmagnetised/weakly magnetised planets (Mars, Venus) alike, and implies a relatively local MM generation in the magnetosheath. The second mechanism, the pickup ion process, is specific to unmagnetised objects with an extended exosphere such as Mars, Venus or comets and creates ring/ring-beam ion velocity distributions which in a high-$\beta$ plasma are unstable to the generation of MMs \cite{Price1989}: it can drive their growth everywhere where the exosphere extends, from the upstream solar wind and in the magnetosheath down to the MPB. In the solar wind, MM-like structures, called MHs or magnetic depressions, have been detected upstream of Mars \cite{madanian_magnetic_2020} and at comets \cite{volwerk_comparison_2014}. This mechanism may become prominent when the bow shock is either quasi-parallel, i.e., less conducing to the rise of anisotropies and more permeable to wave transmission, or weak (for example at comets, and in a lesser measure, at Mars); it may then constitute an important remote source of MMs. Indeed, if the source of the anisotropy is upstream of the shock, MM waves found downstream of the shock but originating upstream of it may consequently have been transported over large distances, almost unchanged, through the quasi-parallel shock. In this latter scenario, if we recall that MH structures in the solar wind are somewhat larger than the MMs observed in the sheath, a damping or attenuation process could be at play when crossing the shock into the magnetosheath, with the trapped particles within the magnetic bottles losing energy. This scenario is reminiscent of the one found by \citeA{plaschke_first_2018} at comet 67P/Churuymov-Gerasimenko for low outgassing rates (and at about the time a nascent bow shock was expected to take take place), with relatively smaller MH structures deep in the coma found to be the end result of upstream larger MHs. 

In our case however, and although contributions from local pickup ions cannot be ruled out entirely, the first mechanism (source in the wake of the quasi-perpendicular shock) seems the likeliest. 
\begin{itemize}
\item First, as we approach closer and closer to the quasi-perpendicular bow shock, the plasma becomes more and more unstable to the generation of MMs, with the detected structures close to the MPB mostly embedded in a plasma marginally stable/unstable to the generation of MMs.
\item Second, the average angle $\alpha_{B-V}$ between the magnetic field direction and the plasma ion velocity is extremely stable around $103\deg\pm5\deg$ (see in Figure\ \ref{fig:MMevent1_contextGeneral}). This implies in velocity phase space that the ion velocity distribution functions (VDF) of newborn pickup ions are closer to a ring distribution, which preferentially contributes to a large increase in $P_\perp$ of these newly picked-up ions as compared to the parallel direction \cite{Price1989}, in agreement with our derived anisotropies. A cursory examination of the VDFs measured by SWIA during the interval (not shown) reveals indeed a non-Maxwellian (or rather non-bi-Maxwellian) behaviour, with free energy available for light particles, such as protons, to drive the growth of instabilities. This may in turn favour the growth of the MM instability over that of the proton cyclotron instability owing to large plasma-$\beta$ and the presence of a small quantity of heavier ions in the plasma \cite{price_numerical_1986}, in keeping with the observations of \citeA{russell_mirror_1987} at comet 1P/Halley. Moreover, because the $\alpha_{B-V}$ angle is large also in the solar wind ($\approx 98\pm5\deg$) as displayed in Figure\ \ref{fig:MMevent1_contextGeneral}, an extra source of anisotropy upstream of the shock may be present, driving MM unstable conditions in those remote locations, with potential mode conversions through the boundary into the magnetosheath.
\item Third, the train of MMs found here shows a variety in shape (dips and peaks), duration and size that point towards different stages in their evolution. 
\item Fourth, the turbulence spectrum during our events and shown in Figure\ \ref{fig:PSD} displays a typical fully developed energy cascade with Kolmogorov-like behaviours, implying that the plasma turbulence has had time to evolve from its generation source.
\item Finally, no magnetic holes are found immediately upstream of the shock, at least in the path of the spacecraft. 
\end{itemize}
All five points above are consistent with the scenario of a nonlocal source region for these MMs likely situated in the bow shock's immediate wake where a large part of the anisotropy grows.

\section{Conclusions}
For the first time, we established the unmistakable presence of a train of linear MMs in the magnetosheath of Mars using in combination MAVEN's magnetic field and high-resolution plasma measurements --- ions and electrons --- in the early part of the mission. Their characteristics have a textbook flair to them: with wave frequencies below the local proton gyrofrequency and lasting approximately $5$ to \qty{10}{s} \cite<as previously found at Venus by>{volwerk_mass-loading_2016}, these magnetic islands follow a clear $B-N_i$ and $B-N_e$ anticorrelation. Trapping distinctly denser and more energetic ions on the inside, they appear to float in a relatively less energetic and more diluted ambient plasma.

These MMs are situated on the dayside in the wake of a highly quasi-perpendicular supercritical shock \cite<as studied by>{Burne2021}, deep in the magnetosheath (\qty{R_{\rm MSO}\sim1.35}{$R_p$}, solar zenith angle \qty{SZA\sim60}{$\deg$}) and close to the transition into the MPB: this is precisely this region where \citeA{ruhunusiri_low-frequency_2015} statistically found waves matching MM characteristics. The upstream solar wind conditions are quiet (\qty{|B|\approx 6.6}{nT}, \qty{V_{sw}\approx 340}{km/s}, \qty{T_i\sim7}{eV}, \qty{T_e\sim12}{eV}, \qty{n_i\sim20}{cm$^{-3}$}), with no CME or other transient solar effects detected at that time. Because of the nature of the shock, magnetic holes created upstream in the solar wind may not easily cross the quasi-perpendicular barrier, which could imply a more local generation mechanism behind the bow shock. Moreover, the magnetic field and plasma signatures of these MMs seem to be in constant evolution: they vary in size (from a few hundreds of kilometres to a few thousands of kilometres) and in peak-to-peak amplitude (from $\approx20$ to \qty{35}{nT} with respect to the background field), possibly implying that these structures are in various stages of evolution. There is also indication that the MMs transit from peaks to dips in a region where the plasma is marginally unstable to MM generation. This could in turn imply that the generator region of these particular MMs, if assumed to be immediately downstream of the shock region, was never probed by the MAVEN spacecraft. Although the quasi-perpendicular shock and its turbulent wake may be the main source of the free energy in the plasma, our results do not exclude two possibilities for the generation of these MMs: (i) local source of anisotropy through pickup ion processes upstream and downstream of the shock, and (ii) MHs forming in the upstream solar wind and being transported through the quasi-perpendicular bow shock into the magnetosheath. However, such a crossing would in practice be difficult and raise questions as to the energy/momentum transfer and wave mode conversion through the shock. 
Finally, the question of their transport from the remote source region behind the shock down to the magnetic pileup boundary where they are detected appears reminiscent of the apparent accumulation of MMs at the Earth's magnetopause \cite{omidi_sources_1994,Erkaev2001,shoji_mirror_2009} and at the Venus induced magnetospheric boundary \cite{volwerk_mirror_2016}.

Further tests of this idea could be done with the use of dedicated plasma flow models in the magnetosheath, as done in \citeA{guicking_low-frequency_2010} for Venus and \citeA{Soucek2015} for Earth, to reconstruct the possible history of MMs along streamlines up to the shock region.
Using the magnetic field-only detection criteria developed in the present work, statistical studies of the location of MMs in Mars' magnetosheath (currently under way) should shed some more light on the place(s) of origin of mirror mode structures, under which conditions they predominantly appear and if the accumulation towards the magnetic pileup boundary is systematic or not.

\appendix
\section{Magnetic Field-Aligned coordinate system}\label{appendix1}

The Magnetic Field-Aligned system (MFA) can be determined by first calculating the ambient magnetic field direction $\v B_\text{ambient}$. In the case of an ion instrument such as MAVEN/SWIA, which takes a finite time in order to scan through angles and energies, one way of estimating $\v B_\text{ambient}$ is to calculate the average magnetic field $\langle{\v B}\rangle$ during a full measurement scan of SWIA. In the SWICA mode during the time interval chosen in this study (11:25-11:30\,UT on 2014-12-25), one measurement takes for instance about $\Delta t = 8$\,s. Another way of estimating $\v B_\text{ambient}$ is to calculate the so-called background field $\v B_\text{bg}$ from a low-pass filtering of $\v B$ as in Section\ \ref{sec:CriteriaB-fieldOnly}.

Assuming the latter for now, the matrix of passage $\v M_\text{MFA}$ between MSO coordinates (base $\{{\bf \hat{X}},{\bf \hat{Y}},{\bf \hat{Z}}\}$) and MFA coordinates (base $\{{\bf \hat{X}^\prime},{\bf \hat{Y}^\prime},{\bf \hat{Z}^\prime}\}$) is usually chosen so that the third axis in the MFA system is aligned with the background $\v B$-field direction, i.e., ${\bf \hat{Z}^\prime} = {\v B_\text{bg}}/|{\v B_\text{bg}}|$. Because in three dimensions there is an infinite number of perpendicular vectors to a given vector, the perpendicular directions are arbitrary: a choice with respect to one perpendicular axis must be made to complete the right-hand rule. One possibility used in the literature is to assume that ${\bf \hat{Y}^\prime}$ is perpendicular to the position vector $\v R$ of the spacecraft \cite{laakso_cluster_2010} so that ${\bf \hat{Y}^\prime} = \v R \times {\bf \hat{Z}^\prime}$. A common choice outside of Earth studies chooses a plane containing the ${\bf \hat{Z}^\prime}$ direction and performs a counterclockwise rotation of $90\deg$ of ${\bf \hat{Z}^\prime}$ so that the rotation matrix between MSO and MFA frames becomes:
\begin{linenomath}
\begin{align}
    {\v M_\text{MFA}} = \{{\bf \hat{X}^\prime},{\bf \hat{Y}^\prime},{\bf \hat{Z}^\prime}\} = \begin{pmatrix}
      0                                                 & \left.{\bf \hat{Z}^\prime}\times{\bf \hat{X}^\prime}\right|_{X} & \hat{Z}^\prime_{X}\\
 \frac{\hat{Z}^\prime_{Z}}{\sqrt{\hat{Z}^{\prime 2}_{Y}+\hat{Z}^{\prime 2}_{Z}}} & \left.{\bf \hat{Z}^\prime}\times{\bf \hat{X}^\prime}\right|_{Y} & \hat{Z}^\prime_{Y}\\
-\frac{\hat{Z}^\prime_{Y}}{\sqrt{\hat{Z}^{\prime 2}_{Y}+\hat{Z}^{\prime 2}_{Z}}} & \left.{\bf \hat{Z}^\prime}\times{\bf \hat{X}^\prime}\right|_{Z} & \hat{Z}^\prime_{Z}
\end{pmatrix}. \label{eq:app:matrixPassageMFA}
\end{align}
\end{linenomath}
This is the convention adopted in the present study. By construction, ${\v M_\text{MFA}}$ is orthogonal, so that the transpose of the matrix is its inverse, i.e., $\v M_\text{MFA}^\intercal = \v M_\text{MFA}^{-1}$, and its determinant obeys $\det\left(\v M_\text{MFA}\right) = 1$.

The instantaneous magnetic field vector in MFA coordinates is simply:
\begin{linenomath}
\begin{align}
	\v B_\text{MFA} = \v M_\text{MFA}^\intercal\ \v B_\text{MSO},
\end{align}
\end{linenomath}
for column vectors.
In the convention above, $B_{||}$ is thus along the $z$ direction and contains most of the signal, whereas the perpendicular directions oscillate closely around zero.

Tests (not shown) were performed in the interval 11:26--11:30\,UT on 2014-12-25 to determine the rotation matrix $\v M_\text{MFA}$ using as ambient field either the \qty{\pm4}{s} averaged $\v B$-field over the SWICA measurement time $\langle{\v B}\rangle$ (which can be argued to be more physical with respect to the ion measurements), or the low-pass filtered field $\v B_\text{bg}$ (used in the $B$-field only approach, a measure of the macroscopic ambient field over a $2$-min span). Both methods yielded very similar results for matrix ${\v M_\text{MFA}}$ in terms of directions and magnitudes. The perpendicular residuals in $\v B_\text{MFA}$ after detrending the magnetic field showed variations of the order of \qty{\pm3}{nT} for a total signal of about \qty{40}{nT} for each method. Consequently, calculations of parallel and perpendicular components of relevant physical quantities such as $\v B_\text{MFA}$, $\overline{\overline{\v P}}_\text{MFA}$ (see \ref{appendix2}) or the mirror mode criterion MMIC of Equation\,(\ref{eq:instabilityCriterion}) were also similar, with only occasional spikes when using $\langle{\v B}\rangle$ linked to the temporal resolution of SWIA (see Figure\ \ref{fig:app:Pressure}). In the main study, the low-pass filtered background field, noted $\v B_\text{bg}$, was used for simplicity and to keep consistent with the MM detection algorithm.

\section{Deriving parallel and perpendicular pressures}\label{appendix2}

As explained in Section\ \ref{sec:criteriaValidation}, the anisotropy in the plasma can be estimated by studying the behaviour of the ion pressure tensor in the Mean Field-Aligned (MFA) coordinate system. The pressure tensor $\overline{\overline{\v P}}$ is a symmetric second-rank tensor with $9$ elements of the form:
\begin{linenomath}
\begin{align}
    \overline{\overline{\v P}} = \begin{pmatrix}
P_{xx} & P_{xy} & P_{xz}\\
P_{xy} & P_{yy} & P_{yz}\\
P_{xz} & P_{yz} & P_{zz}
\end{pmatrix}.
\end{align}
\end{linenomath}
in any $\{\hat{x},\hat{y},\hat{z}\}$ normalised system of coordinates. It is good to recall also for later check-ups that, under any change of coordinate system, three main invariants $\mathcal{I}$ exist for a tensor (Cayley-Hamilton's theorem). For a symmetric matrix, they reduce to:
\begin{linenomath}
\begin{align}
    \mathcal{I}_1 = \tr \overline{\overline{\v P}}  &= P_{xx}+P_{yy}+P_{zz}\\
    \mathcal{I}_2 = \frac{1}{2}\left[\left(\tr \overline{\overline{\v P}}\right)^2 - \tr \overline{\overline{\v P}}^2\right] &= P_{xx}P_{yy}+P_{xx}P_{zz}+P_{yy}P_{zz} - \left(P_{xy}^2+P_{xz}^2+P_{yz}^2\right)\\
    \mathcal{I}_3 = \det \overline{\overline{\v P}} &= P_{xx}P_{yy}P_{zz} + 2P_{xy}P_{xz}P_{yz} - \left(P_{yz}P_{xx}+P_{xz}P_{yy}+P_{xy}P_{zz}\right)
\end{align}
\end{linenomath}
First the pressure tensor, originally expressed in the SWIA instrument coordinate system in eV\,cm$^{-3}$ ($\overline{\overline{\v P}}_\text{swia}$), is rotated into the MSO coordinate system using a rotation matrix from the appropriate SPICE kernel to yield $\overline{\overline{\v P}}_\text{MSO}$. Ideally, in a coordinate system aligned with the mean magnetic field, off-diagonal terms will tend to be small with respect to diagonal terms. Notwithstanding the system of reference, and because the trace of a tensor is invariant with respect to coordinate transforms (rotations), the scalar pressure $p = P_{xx}+P_{yy}+P_{zz}$ remains constant. It follows from this simple statement that when diagonalising the $3\times3$ pressure tensor, the eigenvalues found will stay the same no matter which coordinate system is used.  

To obtain the pressure components parallel and perpendicular to the background or average magnetic field, two methods are usually adopted, the first one based on the direct diagonalisation of the $3\times3$ tensor $\overline{\overline{\v P}}$, the other relatively more involved with a rotation first into the MFA coordinate system and a diagonalisation of the remaining $2\times2$ tensor matrix.

\subsection{Method 1: direct $3\times3$ matrix diagonalisation}
Because of its simplicity and that it can be performed without a supporting magnetometer, this method is usually the one used for onboard moment calculations. In any coordinate system in which the initial pressure tensor is expressed, the pressure tensor can always be diagonalised so that:
\begin{linenomath}
\begin{align}
    \begin{pmatrix}
        P_{\perp 1} & 0          & 0\\
        0           & P_{\perp 2}& 0\\
        0           & 0          & P_{||}
    \end{pmatrix} = \v S^{-1}\ \overline{\overline{\v P}}\ \v S, \label{eq:app:Diag}
\end{align}
\end{linenomath}
with $\v S$ a non-singular matrix containing the right eigenvectors of the system, giving the directions of the principal axes of the diagonalised tensor in the chosen coordinate system. Here the parallel direction to the ambient magnetic field is assumed to be along the $z$ direction, whereas the two other components, $P_{\perp 1}$ and $P_{\perp 2}$, are assumed to be identical in the gyrotropic assumption, by definition $P_{\perp 1}\sim P_{\perp 2}\sim P_\perp$.

In order to forego any assumption on the direction of the ambient magnetic field, one classic way to sort the eigenvalues found in Equation\,(\ref{eq:app:Diag}) is to isolate one that is most different from the two others. This component is by inference the parallel scalar pressure $P_{||}$, whereas the two remaining components are assumed to be the perpendicular pressures. The total perpendicular pressure is then given by $P_\perp = \frac{1}{2} \left(P_{\perp 1}+P_{\perp 2}\right)$.

The situation with negligible off-diagonal terms in the pressure tensor is mostly encountered in the textbook cases of a spherically symmetric distribution function or a gyrotropic one, when the velocity distribution function is cylindrically symmetric about the mean field direction, here assumed along $z$ \cite{paschmann_moments_1998}. However, this situation seldom occurs in practice and determining the parallel and perpendicular pressures from this technique may yield false estimates, sometimes large. 

\subsection{Method 2: alignment to MFA system and $2\times2$ matrix diagonalisation}
When the magnetic field direction is known, this method is arguably more physical than Method $1$, since the pressure tensor is analysed directly in the MFA coordinate system itself instead of an arbitrary coordinate system \cite{swisdak_quantifying_2016}. It thus requires the simultaneous knowledge of the magnetic field direction and amplitude. Calculations can be readily made in instrument coordinates (in which case the $B$-field is transformed into SWIA instrument coordinates) or in MSO coordinates (in which case the pressure tensor is first rotated into the MSO coordinate system). The latter is chosen to keep with the conventions of the main text.

Following the calculation of the matrix of passage from MSO to MFA coordinates (see \ref{appendix1}), the MSO pressure tensor can be rotated into MFA coordinates so that:
\begin{linenomath}
\begin{align}
    \overline{\overline{\v P}}_\text{MFA} = \v M_\text{MFA}^\intercal \left(     \overline{\overline{\v P}}_\text{MSO}\ \v M_\text{MFA} \right) =  \begin{pmatrix}
P_{XX}^\prime & P_{XY}^\prime & P_{XZ}^\prime\\
P_{XY}^\prime & P_{YY}^\prime & P_{YZ}^\prime\\
P_{XZ}^\prime & P_{YZ}^\prime & P_{ZZ}^\prime
\end{pmatrix}
\end{align}
\end{linenomath}
is also a symmetric tensor.

By definition, $P_{ZZ}^\prime = P_{||}$, whereas $P_{XY}^\prime$, $P_{YZ}^\prime$ and $P_{XZ}^\prime$ are non-zero off-diagonal elements indicative of shear stresses, i.e., the flow of momentum in the $x$ (respectively, $y$) direction by motion of the plasma in the $y$ ($z$) direction. Keeping the third column of the tensor aside, and since the determination of perpendicular directions is always arbitrary (see \ref{appendix1}), the two perpendicular directions are simply obtained by diagonalising the remaining $2\times2$ matrix (which is always diagonalisable, by construction):
\begin{linenomath}
\begin{align}
 \begin{pmatrix}
P_{\perp 1} & 0\\
0 & P_{\perp 2}
\end{pmatrix} = \v S^{\prime -1}\ \begin{pmatrix}
P_{XX}^\prime & P_{XY}^\prime\\
P_{XY}^\prime & P_{YY}^\prime
\end{pmatrix}\ \v S^\prime,
\end{align}
\end{linenomath}
where $\v S^\prime$ is a singular matrix containing the right eigenvectors of the system. By convention, the perpendicular component eigenvalues are sorted by increasing value.

Additionally, a measure of the apparent non-gyrotropy can be extracted from the off-diagonal elements, assuming that $P_{\perp} = \left(P_{\perp1}+P_{\perp2}\right)/2$ as in \citeA{swisdak_quantifying_2016}:
\begin{linenomath}
\begin{align}
    \zeta = \frac{P_{XY}^{\prime2}+P_{XZ}^{\prime2}+ P_{YZ}^{\prime2}}{P_\perp^2+2P_{||}P_\perp} = 1 - \frac{4\mathcal{I}_2}{\left(\mathcal{I}_1 - P_{||}\right)\left(\mathcal{I}_1+3P_{||}\right)} \label{eq:app:gyrotropy}
\end{align}
\end{linenomath}
The so-called gyrotropy index $\zeta$ varies between $0$ (gyrotropic tensor) and $1$ (maximum departure from gyrotropy).

\begin{figure}
		\noindent\includegraphics[width=\textwidth]{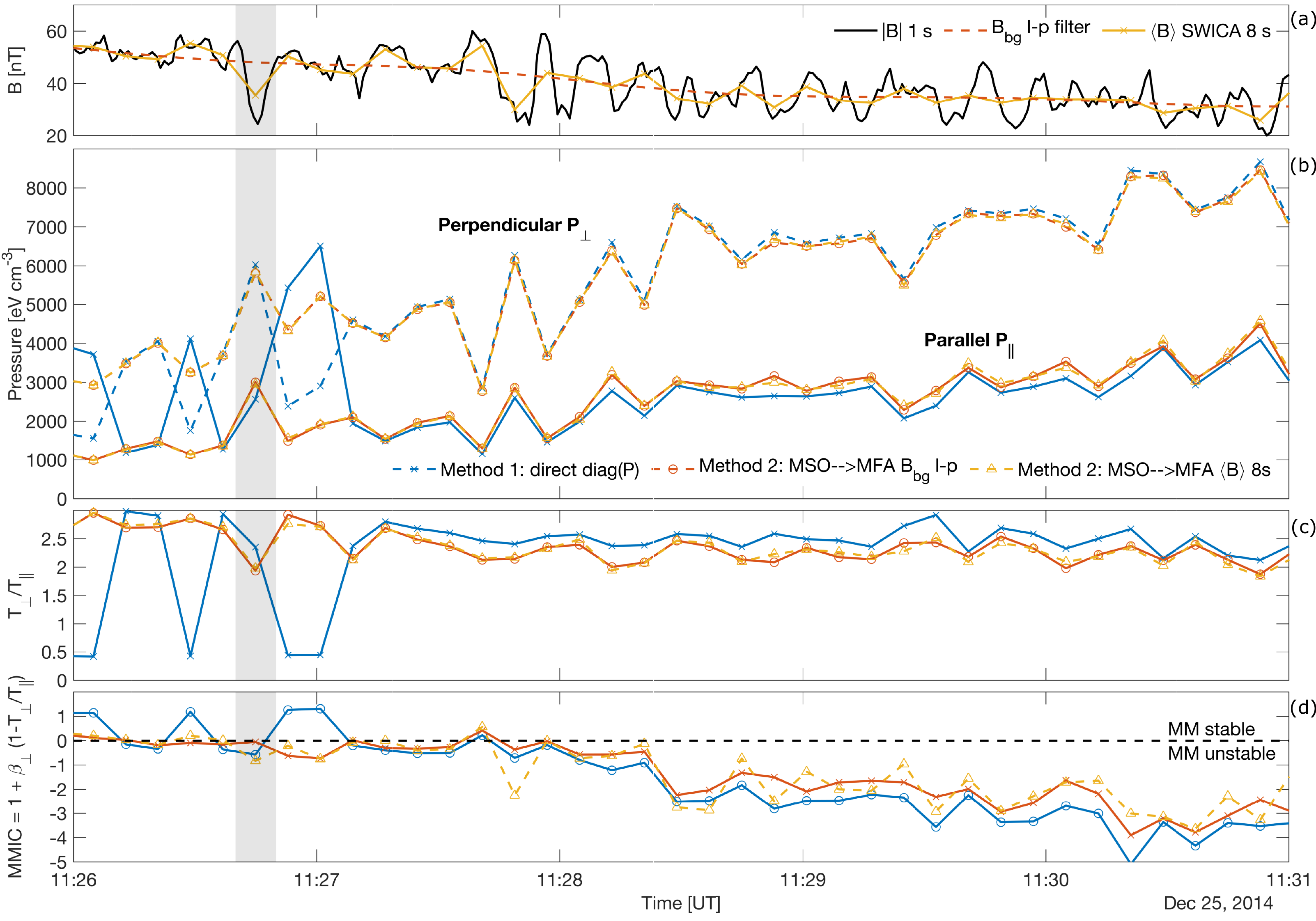}
		\caption{Comparison of retrievals for the pressure tensor parallel and perpendicular components. (a) Magnetic field $|\v B|$ from MAVEN/MAG at \qty{1}{s} resolution, low-pass-Butterworth filtered field $|\v B_\text{bg}|$ as in Section\ \ref{sec:CriteriaB-fieldOnly}, and average field $\langle |\v B|\rangle$  over SWICA's mode resolution (\qty{8}{s}, centered), during the interval 11:26--11:30\,UT on 2014-12-25. 
		(b) Parallel $P_{||}$ (continuous lines) and total perpendicular ion pressures (dashed line) $P_{\perp} = \frac{1}{2} \left(P_{\perp 1} + P_{\perp 2}\right)$, as retrieved from MAVEN/SWIA (SWICA mode).
		(c) Temperature (or pressure) anisotropy. 
		(d) Mirror mode instability criterion (MMIC), with the zero line separating MM-unstable and MM-stable conditions. Throughout panels $2$ to $4$, method $1$ (direct diagonalisation) is in blue whereas method $2$ is in orange and yellow (tensor expressed in MFA, using either $\v B_\text{bg}$ or $\langle \v B\rangle$, respectively, to estimate the ambient $B$-field direction). The grey-shaded zone represents the clearest MM structure in this interval as detected by a combination of plasma and magnetic field measurements.
		}
		\label{fig:app:Pressure}
\end{figure}

\subsection{Comparison of methods}
A comparison of the two methods is presented in Figure\ \ref{fig:app:Pressure}, with Method $1$ (direct diagonalisation) shown in panels (b-d) as a blue line, whereas Method $2$ results are given in orange (and, for comparison, yellow, when using SWIA's mode scanning resolution of \qty{8}{s}). Although local differences are clearly seen, the two methods agree rather well on average. At the time when a clear MM structure is detected (grey zone), $P_\perp \sim 2.5 P_{||}$ for both methods.
However, at the beginning of the interval, Method $1$ exhibits abrupt variations of a factor $3-4$ over a time scale comparable with the SWICA scanning temporal resolution (\qty{\sim8}{s}): this is highly suspicious and linked to the rather arbitrary sorting of the eigenvalues, with the two closest eigenvalues assumed to be the two perpendicular components, in keeping with the gyrotropic assumption. Method $2$, on the contrary, provides a more gradual and smoother evolution of the pressures, both parallel and perpendicular, as well as the temperature ratios. This appears more consistent with the measurements themselves and the physics at play. Except from a few less gradual point-to-point variations, a similar conclusion can be made when using Method $2$ and estimating the ambient field direction from the mean field $\langle \v B\rangle$ over the measurement time of SWIA (yellow) instead of using the macroscopic low-pass filtered $\v B_\text{bg}$.

Because of the arbitrary way the parallel component is chosen in the diagonalised pressure tensor in Method $1$, only Method $2$ should be safely used in this particular case especially at the beginning of the interval, which has repercussions on any other quantities derived from the ion pressure tensor: $T_{||}$ and $T_\perp$, plasma-$\beta_{||}$ and $\beta_\perp$. Ultimately, this may alter the interpretation of the MMIC of Equation\,(\ref{eq:instabilityCriterion}). This is shown in Figure\ \ref{fig:app:Pressure}d, where Method $1$ would imply sharp oscillations around the MM-stable line at the beginning of the interval, which is misleading. In contrast, Method $2$ predicts marginally MM-stable conditions (MMIC$\approx0$) in this time span. After 11:28\,UT, both methods agree rather well.

It is important to note here that the validity of any of those methods depends on the quality of the ion velocity distribution measured in the first place, and the field of view (FOV) of the instrument \cite{halekas_flows_2017}. If the direction of the magnetic field for example lies in the blind sectors of the plasma instrument due its limited FOV,
the parallel estimate of the pressure tensor will become difficult to assess and likely underestimated. Likewise, if the instrument does scan through the direction where the magnetic field points, the parallel direction and only one perpendicular direction (corresponding to the largest eigenvalue) will be well-defined. Assumptions on either $P_{||}$ or $P_\perp$ must thus be made depending on the case. For SWIA, the coarse mode usually adopted in the magnetosheath has an angular FOV of $360\deg\times90\deg$ (SWICA mode), which, despite its broad coverage in comparison to other more FOV-limited modes, needs to be checked against the magnetic field direction for each event, separately.

In conclusion, Method $2$, being more physical, should always be preferred over Method $1$ when magnetic field measurements are available, unless the velocity distribution function is shown to be spherically or cylindrically symmetric. Vigilant care in the interpretation is strongly recommended in all cases.

%
%
%
%
%
%
%
%

\acknowledgments
C. Simon Wedlund, M. Volwerk and C. Möstl thank the Austrian Science Fund (FWF): P32035-N36, P31659-N27, P31521-N27. Parts of this work for the observations obtained with the SWEA
instrument are supported by the French space agency CNES. The authors also acknowledge Emmanuel Penou for help and access to the CLWeb viewing and analysis software (v16.09) from IRAP/Observatoire Midi-Pyrénées.
CSW thanks Arnaud Beth (Umeå University, Sweden) for useful comments, M. Simon Wedlund for patient and constructive discussions, and acknowledges Yair Altman for developing and maintaining the Matlab package ``export\_fig'' for figure pdf exports.
The authors thank ISSI and the ISSI international team 517 ``Towards a Unifying Model for Magnetic Depressions in Space Plasmas'' led by MV, and team 499 ``Similarities and Differences in the Plasma at Comets and Mars'' led by Charlotte Götz, for facilitating research between team members during Covid times.
Throughout the manuscript all inverse tangents were calculated using the $\rm atan2$ function in order to resolve all four angular quadrants. 

{\bf Data Availability Statement.} The calibrated MAVEN/MAG, SWIA and SWEA datasets are freely available from the NASA Planetary Data System (PDS), respectively at \url{https://doi.org/10.17189/1414178}, \url{https://doi.org/10.17189/1414182} and \url{https://doi.org/10.17189/1414181}.


%
%

\bibliography{bibliography}

%
%
%
%
%

\end{document}